\documentclass[twocolumn]{aastex631}
\usepackage[nolist]{acronym}

\newcommand\Msun{\; {\rm M}_{\odot}}
\newcommand\kms{\; {\rm km}\;{\rm s}^{-1}}

\newcommand\kpc{\;{\rm kpc}}
\newcommand\freq{\kms\kpc^{-1}}

\newcommand\Gyr{\;{\rm Gyr}}

\newcommand\QT{Q_{T}}
\newcommand\QTmin{{Q_{T,\text{min}}}}
\newcommand\Xbar{{X_\text{bar}}}
\newcommand\Qbar{{{Q_{T,\rm bar}}}}
\defcitealias{jnk23}{JK23}

\usepackage{amsmath}
\usepackage{graphicx}
\usepackage{soul}
\usepackage{multirow}
\usepackage[nameinlink,capitalise,noabbrev]{cleveref}

\shorttitle{Bar Formation Condition}
\shortauthors{JANG, KIM, \& Lee}

\begin{document}

\title{Conditions for Bar Formation in Bulgeless Disk Galaxies}


\author[0000-0002-7202-4373]{Dajeong Jang} 
\affiliation{Department of Physics $\&$ Astronomy, Seoul National University, Seoul 08826, Republic of Korea}

\author[0000-0003-4625-229X]{Woong-Tae Kim}
\affiliation{Department of Physics $\&$ Astronomy, Seoul National University, Seoul 08826, Republic of Korea}
\affiliation{SNU Astronomy Research Center, Seoul National University, Seoul 08826, Republic of Korea}
\email{unitree@snu.ac.kr}

\author[0000-0003-2779-6793]{Yun Hee Lee}
\affiliation{Department of Astronomy and Atmospheric Sciences, Kyungpook National University, Daegu 41566, Republic of Korea}

\begin{acronym}
    \acro{BPS}{boxy peanut-shaped}
    \acro{ILR}{inner Lindblad  resonance}
\end{acronym}

\begin{abstract}
While bars are commonly observed in disk galaxies, the precise conditions governing their formation remain incompletely understood. To investigate these conditions, we perform a suite of $N$-body simulations of bulgeless disk galaxies with stellar masses in the range $10^{9} \leq M_d \leq 10^{11}\Msun$. Our galaxy models are constructed based on the observed properties of nearby barred galaxies from the S$^4$G survey, and we systematically vary the halo scale radius to isolate its dynamical influence. Bars in our simulations form via repeated swing amplifications of disk perturbations, sustained by feedback loops. The amplification factor $\Gamma$ depends on both the Toomre stability parameter $\QT$ and the dimensionless wavelength $X$. Based on our simulation results, we propose a two-parameter bar formation criterion, $Q_T + 0.4(X - 1.4)^2 \leq 1.8$, corresponding to $\Gamma = 10$, which better captures the onset of bar formation than traditional one-parameter conditions. Bars in low-mass galaxies tend to be shorter and weaker, and are more susceptible to disruption by outer spiral arms. In contrast, bars in high-mass galaxies are longer, stronger, and more resilient to spiral interference. Bars in low-mass galaxies undergo only slight vertical thickening over time, whereas those in high-mass galaxies thicken rapidly via buckling instability.
\end{abstract}


\section{Introduction} \label{sec:intro}

Optical and near-infrared surveys reveal that bar structures are prevalent in disk galaxies, with over $\sim60\%$ of local galaxies possessing either weak or strong bars \citep{de63,sw93,kna00, whyte02,lauri04,marijogee07,menendez07,agu09,mendez12,buta15,diaz16,diaz19,lee19}. 
Advances in observational techniques, particularly recent results from JWST and other surveys, have enabled the detection of bar structures at redshifts as high as  $z \sim 4$ \citep{guo23,huang23,conte24,amvro25,guo25,geron25}.
Bars are also found across a wide range of galaxy types including dwarf galaxies and low-surface-brightness galaxies \citep{janz12,banerjee13,sodi17,peters19,michea21,cuomo24} and span a broad range of stellar masses \citep{erwin18,erwin19,erwin24}. Since bars play a key role in driving secular evolution and redistributing angular momentum within galaxies, it is crucial to understand how bars form and evolve in diverse galactic environments and under varying structural and dynamical conditions.

Theoretically, bars are thought to arise from gravitational instability in rotationally supported stellar disks \citep{toomre64}, where non-axisymmetric perturbations grow through swing amplification of leading waves into trailing ones, transforming otherwise nearly circular stellar orbits into elongated $x_1$ orbits \citep[e.g.,][]{sell14}. The strength of swing amplification in a disk with surface density $\Sigma$ depends on two dimensionless parameters \citep[e.g.,][]{bnt08}:
\begin{equation}\label{eq:X}
Q_T \equiv \frac{\kappa \sigma_R}{3.36G\Sigma }, \qquad
X \equiv \frac{\kappa^2 R}{2\pi G \Sigma m},
\end{equation}
where $\kappa$ is the epicycle frequency, $\sigma_R$ the radial velocity dispersion, and $m$ the azimuthal mode number. Note that $Q_T$ is the Toomre stability parameter, and $X$ characterizes the azimuthal wavelength. Calculations show that the amplification factor increases with decreasing $Q_T$ and peaks near $X \sim 1.4$ \citep{GLB65,JT66,toomre81,michi16}.
If a single swing amplification is insufficient to induce bar formation, spiral waves may undergo repeated amplification through feedback loops that convert trailing waves into leading ones.

The presence of a fixed dark halo can strongly inhibit bar formation by increasing the epicycle frequency \citep{onp73,hohl76}, whereas a live halo facilitates bar growth through angular momentum exchange \citep{ath02,mar06,col18,col19a,col19b}. A number of halo properties have been shown to influence bar formation and evolution, including the axial ratio \citep{ath02,ath13}, spin \citep{sn13,long14,col18,col21,kns22,li23arx,jnk24}, and concentration \citep{athmis02,ath02}.

Bars embedded in triaxial halos tend to form earlier than those in spherical halos \citep{ath13}, although halo triaxiality can also suppress bar growth through nonlinear interactions between the stellar bar and halo major axes. \citet{col18} further demonstrated that bar formation proceeds most efficiently in spherical halos, while it is delayed in prolate halos. Bars also tend to develop more strongly in galaxies with a more centrally concentrated halo \citep{athmis02,ath02}, although the density profiles of halos remain observationally uncertain, as highlighted by the core–cusp problem \citep{flores94,moore94}. A classical bulge may likewise inhibit bar formation by increasing the epicycle frequency, although a live bulge can also promote bar growth by absorbing angular momentum from the disk \citep{sell80,se18,kd18,fujii18}.

Although previous numerical studies have provided valuable insights into the formation and evolution of bars, there remains no consensus on the physical conditions required to produce a bar in disk galaxies. In a pioneering study employing fixed halos, \citet{onp73} found that bar formation occurs when 
\begin{equation}\label{eq:top}
  t_\text{OP} \equiv T/|W| > 0.14,
\end{equation} 
where $T$ and $W$ denote the total rotational and gravitational potential energies of the galaxy, respectively. Using two-dimensional models with a fixed halo, \cite{efsta82} proposed a criterion for bar formation, known as the ELN-criterion, given by 
\begin{equation}\label{e:eELN}
    \epsilon_\text{ELN}  \equiv \frac{V_{\rm max}}{(GM_{d}/R_{d})^{1/2}} < 1.1,
\end{equation}
where $V_{\rm max}$ is the maximum rotational velocity, and $M_d$ and $R_d$ are the mass and scale radius of the disk, respectively. For galaxies with classical bulges, \citet{kd18} proposed that bar formation requires $\mathcal{F}_\text{KD} \equiv GM_b/R_dV_\text{tot}^2 < 0.35$, where $M_b$ is the bulge mass and $V_\text{tot}$ is the rotational velocity at $R=R_d$. Alternatively, \citet{se18} suggested the criterion $\mathcal{D}_\text{SE} \equiv \left<\rho_b\right>/\left<\rho_d\right> < 1/\sqrt{10}$,
where $\left<\rho_b\right>$ and $\left<\rho_d\right>$ are the mean bulge and disk densities within the bulge half-mass radius, respectively. 

Because swing amplification is governed by the two independent parameters $\QT$ and $X$, the single-parameter criteria mentioned above, typically formulated in terms of energy ratios, are overly simplistic and may not adequately capture the efficiency of the process. To illustrate this, \citet{jnk23} performed $N$-body simulations of Milky Way–like galaxy models, systematically varying the mass and compactness of the classical bulge as well as the halo concentration. Their results indicate that bar formation in these models is more accurately characterized by the following two-parameter criterion:
\begin{equation}\label{e:cri}
\left(\frac{\QTmin}{1.2}\right)^2 + \left(\frac{\text{CMC}}{0.05}\right)^2 < 1,
\end{equation}
where $\QTmin$ is the minimum value of the stability parameter and CMC denotes the central mass concentration. This condition is physically motivated, as bar formation depends on repeated cycles of swing amplification, which operate more effectively in disks with lower $\QTmin$ and CMC \citep{sell80,toomre81,bnt08}.

While \cref{e:cri} accounts for the numerical results of bar formation reasonably well, it is primarily applicable to relatively massive galaxies, such as the Milky Way. However, observations show that galaxies span a broad range of sizes and masses, with Milky Way–like systems representing only a fraction of this diverse population.  For example, \citet{diaz16} found that the Milky Way belongs to a relatively massive subgroup comprising only about 10\% of nearby barred galaxies. Based on the \emph{Spitzer} Survey of Stellar Structure in Galaxies (S$^4$G), \citet{erwin18} found that the bar fraction, $f_\text{bar}$, peaks at $M_* \sim 10^{9.7}\Msun$ and declines at higher stellar masses, reaching $f_\text{bar} \sim 0.5$ for Milky Way–like galaxies with $M_* \sim 10^{10.7}\Msun$. Using the Sloan Digital Sky Survey database, \citet{oh12} reported that $f_\text{bar} < 0.4$ at $M_* = 10^{10.7}\Msun$, with $f_\text{bar}$ increasing with $M_*$.
This raises the question of whether the bar formation criteria discussed above are applicable across the full spectrum of galaxy masses.

In this paper, we revisit the issue of bar formation in disk galaxies by systematically varying their mass and size. We construct a suite of isolated galaxy models that span a broad range of disk masses and scale lengths, designed to reflect the observed diversity among barred systems reported by \citet{diaz16}. This allows us to explore how fundamental structural parameters influence bar formation and evolution. In the present study, we restrict our analysis to galaxies without a classical bulge in order to isolate the effects of disk and halo properties.
Using high-resolution $N$-body simulations, we examine the conditions under which bars form, characterizing them in terms of the dimensionless $Q_T$ and $X$ parameters that govern the efficiency of swing amplification. The numerical results are then compared with analytic predictions. In addition, we investigate how the resulting bar properties, such as strength, length, and pattern speed, depend on galaxy mass.

This paper is organized as follows. In \autoref{sec:modelnmethod}, we describe the galaxy models and numerical methods employed in our study. \autoref{sec:results} presents the results of the $N$-body simulations, focusing on bar formation and evolution, as well as interactions between the bar and outer spiral arms. We also explore how bar length, pattern speed, and the onset of buckling instability vary with galaxy mass. In \autoref{sec:condition}, we interpret the conditions for bar formation in the context of the analytic theory of swing amplification. In \autoref{sec:discussion}, we interpret our numerical results in terms of $t_\text{OP}$ and $\epsilon_\text{ELN}$, compare the bar strengths and lengths between simulations and observations, and discuss the criteria for buckling instability. Finally, we summarize our findings in \autoref{sec:summary}.

\section{Galaxy Model and Method}\label{sec:modelnmethod}

In this section, we describe the galaxy parameters chosen to represent observed nearby barred galaxies. We also present the initial galaxy models and outline the numerical methods employed in our simulations.

\begin{figure}[t]
\centering
\epsscale{1.0} \plotone{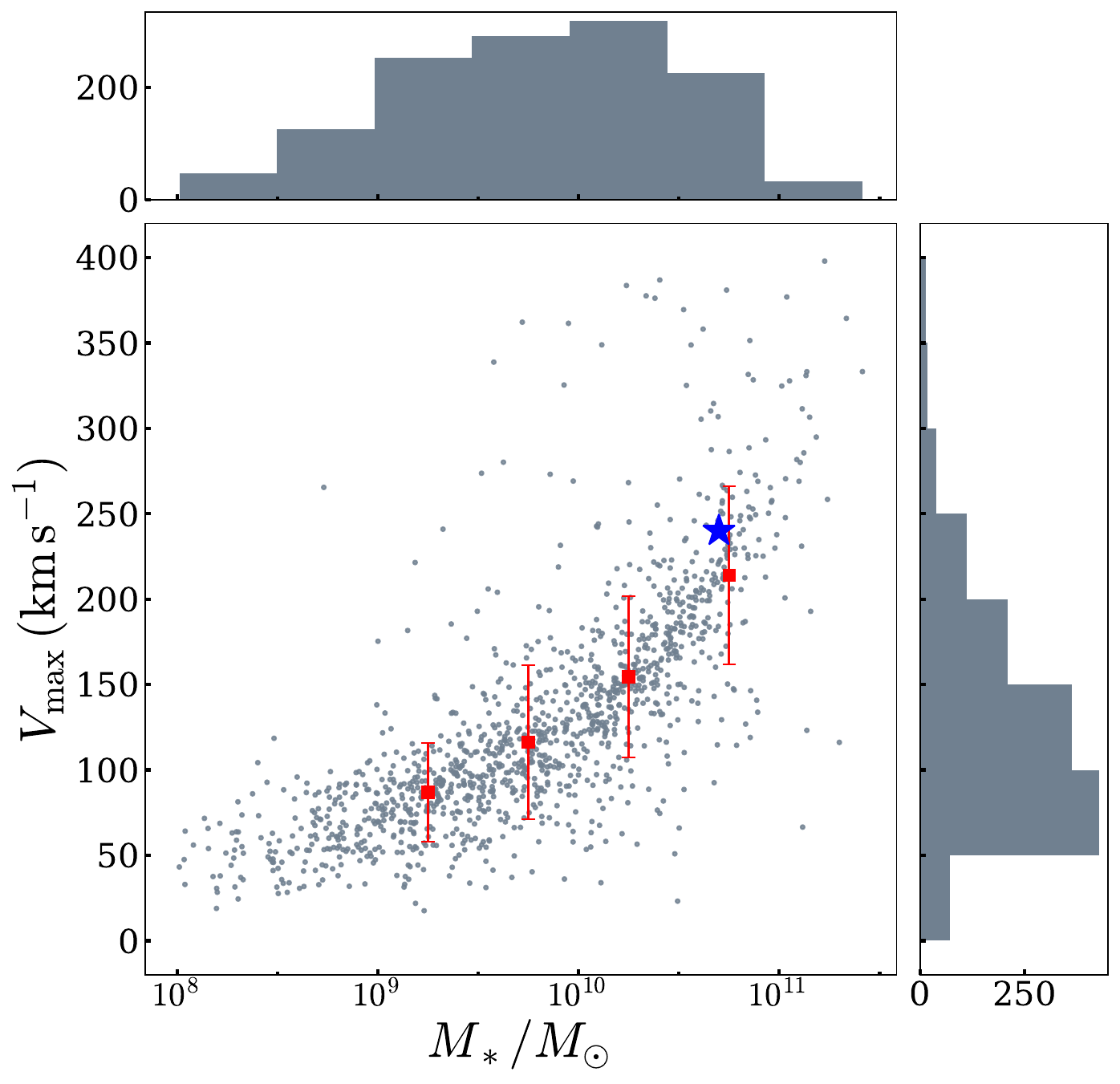}
\caption{Distribution of S$^{4}$G barred galaxies from \citet{diaz16} in the plane defined by the maximum \ion{H}{1} circular velocity, $V_\text{max}$, and stellar mass, $M_*$. The top and right panels draw the histograms of $M_*$ and $V_\text{max}$, respectively. The blue star symbol marks the location of the Milky Way, while the red squares with error bars indicate the mean and standard deviation within each of the four stellar mass bins. 
\label{fig:sample}}
\end{figure}

\begin{deluxetable*}{clccccccccc}
\tablecaption{Model  parameters\label{tbl:model}}
\tablenum{1}
\tablehead{
\colhead{Group} &
\colhead{Name\tablenotemark{a}} & 
\colhead{$M_d$} & 
\colhead{$R_d$} &
\colhead{$z_d$} &
\colhead{$M_h$} &
\colhead{$a_h$\tablenotemark{b}} &
\colhead{$V_{\text{max}}$} &
\colhead{$\Qbar$} &
\colhead{$\Xbar$} &
\colhead{$t_\text{bar}$} \\
\colhead{}&
\colhead{}&
\colhead{($10^{10}\Msun$)} & 
\colhead{(kpc)}  &
\colhead{(kpc)}  &
\colhead{($10^{10}\Msun$)} &
\colhead{(kpc)}  &
\colhead{(${\rm km}\;{\rm s}^{-1}$)}  &
\colhead{}&
\colhead{}&
\colhead{(Gyr)}\\
\colhead{(1)} & \colhead{(2)} & \colhead{(3)} & \colhead{(4)} & \colhead{(5)} & \colhead{(6)} & \colhead{(7)} & \colhead{(8)} & \colhead{(9)} & \colhead{(10)} & \colhead{(11)} 
}
\startdata
\multirow{6}{1em}{\rotatebox[]{90}{Group 1}} &
\texttt{G1A25}   &\multirow{6}{2em}{0.178} & \multirow{6}{2em}{2.0}  & \multirow{6}{2em}{0.25}& \multirow{6}{2em}{17.3}  & 25 &88& 1.8& 3.1& \nodata \\
&\texttt{G1A28}   &   &  &  &  & 28 &83& 1.6 & 2.6 & \nodata\\
&\texttt{G1A31}   &   &  &  &  & 31 &78& 1.5 & 2.4 & \nodata\\
&\texttt{G1A34}   &   &  &  &  & 34 &74& 1.4 & 2.2 & 9.4\\
&\texttt{G1A37}   &   &  &  &  & 37 &71& 1.3 & 1.9 & 5.3\\
&\texttt{G1A40*}  &   &  &  &  & 40 &67& 1.3 & 1.8 & 5.3\\
\hline
\multirow{5}{1em}{\rotatebox[]{90}{Group 2}} &
\texttt{G2A28}   & \multirow{5}{2em}{0.562}  &\multirow{5}{2em}{2.5} & \multirow{5}{2em}{0.30} &\multirow{5}{2em}{32.0} & 28 &115& 1.5 &  2.3 & \nodata\\
&\texttt{G2A30}   &   &  &  &  & 30 &111& 1.4 & 2.1 & 8.5\\
&\texttt{G2A32}   &   &  &  &  & 32 &107& 1.4 & 2.0 & 6.9\\
&\texttt{G2A34}   &   &  &  &  & 34 &104& 1.3 & 2.0 & 5.8\\
&\texttt{G2A36*}  &   &  &  &  & 36 &100& 1.3 & 1.8 & 4.5\\
\hline
\multirow{6}{1em}{\rotatebox[]{90}{Group 3}} &
\texttt{G3A25}  & \multirow{6}{2em}{1.778}   &\multirow{6}{2em}{3.0} & \multirow{6}{2em}{0.50} &\multirow{6}{2em}{66.4}& 25 &180& 1.9 & 2.2 & \nodata\\
&\texttt{G3A27}  &   &  &  &  & 27 &174& 1.8& 2.1 & \nodata\\
&\texttt{G3A29}  &   &  &  &  & 29 &168& 1.7& 1.9 & 7.9\\
&\texttt{G3A31}  &   &  &  &  & 31 &162& 1.6& 1.8 & 6.6\\
&\texttt{G3A33}  &   &  &  &  & 33 &157& 1.6& 1.6 & 4.6\\
&\texttt{G3A35*} &   &  &  &  & 35 &152& 1.5& 1.6 & 4.1\\
\hline
\multirow{6}{1em}{\rotatebox[]{90}{Group 4}} &
\texttt{G4A35}  & \multirow{6}{2em}{5.623}  &\multirow{6}{2em}{4.0} & \multirow{6}{2em}{0.80} &\multirow{6}{2em}{219.7} & 35 &277& 2.1 & 2.0 & \nodata\\
&\texttt{G4A40}  &  &  &  &  & 40 &258& 1.9 & 1.6 & \nodata\\
&\texttt{G4A45}  &  &  &  &  & 45 &243& 1.8 & 1.3 & 6.7\\
&\texttt{G4A50}  &  &  &  &  & 50 &229& 1.7 & 1.2 & 5.7\\
&\texttt{G4A55}  &  &  &  &  & 55 &218& 1.6 & 1.2 & 3.7\\
&\texttt{G4A60*} &  &  &  &  & 60 &209& 1.5 & 1.0 & 3.5\\
\enddata 
\tablecomments{Column (1): group name; Column (2): model name; Column (3): disk mass; Column (4): disk scale radius; Column (5): disk scale height; Column (6): halo mass; Column (7): halo scale radius; Column (8): total rotational velocity at $R=20 \kpc$; Column (9): averaged value of the Toomre stability parameter; Column (10): averaged value of the dimensionless azimuthal wavelength; Column (11): epoch of bar formation
}
\tablenotemark{a}{The models marked with an asterisk represent the fiducial models.}

\tablenotemark{b}{The range of $a_h$ is set not to cover the observed values of $V_\text{max}$ in \cref{fig:sample}, but to span both stable and unstable models.}
\end{deluxetable*}

\subsection{Galaxy Samples}\label{sec:sample}

As a basis for constructing our galaxy models, we utilize data from the S$^4$G survey \citep{sheth10}. \citet{diaz16} selected a sample of barred galaxies with disk inclinations $i < 65^{\circ}$, based on the morphological classification scheme of \citet{buta15}, and analyzed their stellar mass distributions by converting the 3.6$\;\mu$m flux using a fixed mass-to-light ratio of $M_*/L_* = 0.53$. For each galaxy in their sample, they inferred the maximum circular velocity, $V_\text{max}$, from the \ion{H}{1} line width provided by the Cosmic Flows project \citep{Courtois09, Courtois11}.

\autoref{fig:sample} plots the relation between $V_\text{max}$ and stellar mass $M_*$, along with their respective histograms, for the sample of 1345 disk galaxies compiled by \citet{diaz16}. Galaxies with greater mass typically rotate more rapidly. In this study, we assume that the entire stellar mass resides in the disk, neglecting any contribution from a bulge component. The blue star symbol marks the position of the Milky Way, highlighting its status as a relatively massive and rapidly rotating galaxy. Most of the galaxies lie within the mass range $10^{9} \Msun \leq M_* \leq 10^{11} \Msun$. We divide this range into four groups: $\log (M_*/{\rm M}_\odot) = [9.0, 9.5]$, $[9.5, 10.0]$, $[10.0, 10.5]$, and $[10.5, 11.0]$. The red circles indicate the median stellar mass and average maximum circular velocities $\overline{V}_\text{max}=87, 116, 155$, and $214\kms$ for Groups 1 through 4, respectively,  with error bars representing the corresponding standard deviations $\sigma_{V_\text{max}}=29, 45, 47$, and $52\kms$. In each group, we fix the disk mass and scale radius, and determine the halo mass using the empirical halo-to-stellar mass relation \citep[see, e.g.,][]{moster10,guo10,leau12,diaz16}. To explore the effect of $V_\text{max}$, we systematically vary the halo scale radius, which modulates the shape of the rotation curve while preserving the total halo mass.

\begin{figure*}[t]
\centering
\epsscale{0.8} \plotone{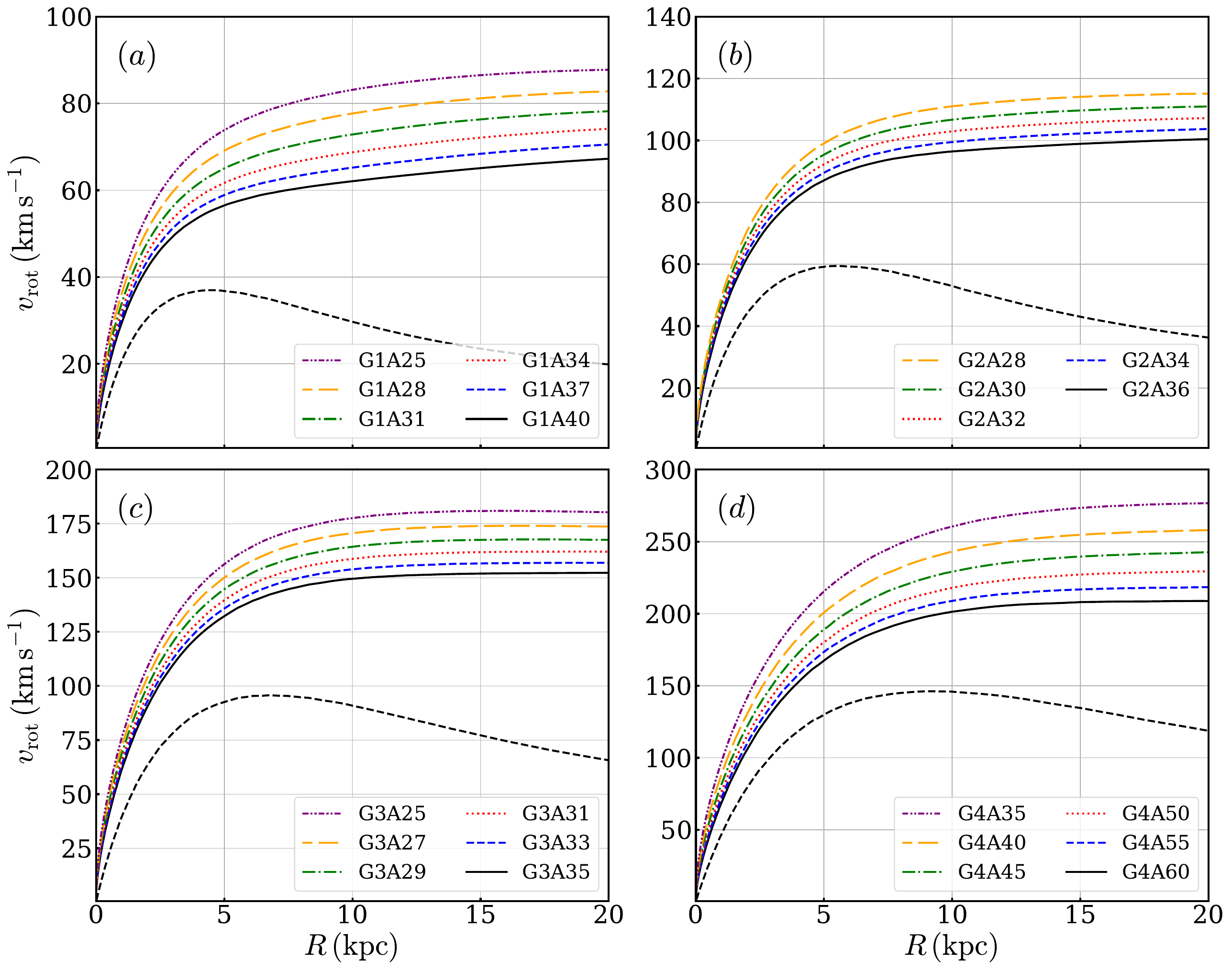}
\caption{Initial distributions of the circular velocity $v_\text{rot}$ as a function of radius, shown as solid lines for the models in $(a)$ Group 1, $(b)$ Group 2, $(c)$ Group 3, and $(d)$ Group 4. The dashed line in each panel indicates the contribution from the disk component. 
\label{fig:vrot}}
\end{figure*}

\subsection{Initial Galaxy Models}

Our galaxy models comprise a stellar disk, a dark matter halo, and a central supermassive black hole. The stellar disk follows a standard exponential-sech$^2$ vertical density profile:
\begin{equation}
\rho_d(R,z) = \frac{M_d}{4\pi z_dR_d^2} \exp\left( -\frac{R}{R_d}\right)
{\rm sech}^2\left( \frac{z}{z_d}\right),
\end{equation}
where $R$ is the cylindrical radius, $R_d$ is the disk scale radius, $z_d$ is the disk scale height, and $M_d$ is the disk mass.  For the four groups, we fix the disk mass at $\log_{10}(M_{d}/{\rm M}_{\odot}) = 9.25$, 9.75, 10.25, and 10.75, respectively, and adopt corresponding scale lengths of $R_d = 2.0$, 2.5, 3.0, and $4.0\kpc$, and scale heights of $z_d = 0.25$, 0.3, 0.5, and $0.8\kpc$, reflecting the mean values of the samples in each group.

The dark matter halo follows \cite{hernquist90} density profile:
\begin{equation}
\rho(r) = \frac{M_h}{2\pi}\frac{a_h}{r(r+a_h)^3},
\end{equation}
where $r=(R^2+z^2)^{1/2}$ is the spherical radius, and $M_h$ and $a_h$ are the mass and the scale radius of the halo. Using the halo-to-stellar mass relation from \citet{diaz16}, we constrain the halo mass to $M_h/(10^{10}\Msun) = 17.3$, 32.0, 66.4, and 219.7 for the four galaxy groups, respectively.  To account for uncertainties in the observed values of $V_\text{max}$ and to identify the conditions for bar formation, we vary $a_h$ within each group to span the range between bar-stable and bar-unstable models. A central black hole is placed at the center of each galaxy, with masses of $M_\text{BH}/(10^6 \Msun) = 0.18$, 0.65, 2.33, and $8.37$, determined using the black hole-to-stellar mass relation from \citet{cho24}. 

\autoref{tbl:model} summarizes the model names, initial galaxy parameters, and simulation outcomes. Columns (1) and (2) list the group and model names, respectively. The prefix \texttt{G} followed by a number designates the group, while the infix \texttt{A} followed by a number indicates the halo scale radius.
We designate the model with the largest halo scale radius within each group as the fiducial model (marked with an asterisk), which is found to be the most unstable to bar formation.  Columns (3)--(5) provide the mass, scale radius, and scale height of the disk, respectively, while Columns (6) and (7) list the mass and scale radius of the halo. Column (8) gives the maximum rotational velocity.\footnote{The adopted range of $V_\text{max}$ is not intended to reproduce the observed distribution shown in \cref{fig:sample}, but rather to encompass both stable and unstable models.} Columns (9) and (10) list $\Qbar$ and $\Xbar$ at $t = 0$, which are the radial averages of $Q_T$ and $X$ with $m=2$ over the range $2\kpc \leq R \leq R_{\QTmin}$, respectively, with $R_{\QTmin}$ denoting the radius at which $\QT$ attains its minimum.  Finally, Column (10) lists the bar formation time $t_\text{bar}$. Models without a reported value of $t_\text{bar}$ remain stable and do not form bars within $10\Gyr$. 

\Cref{fig:vrot} compares the initial radial profiles of the circular velocity, $v_c(R) = \big(R\partial \Phi/\partial R\big)^{1/2}$, computed from the gravitational potential $\Phi$, for all 23 models, displayed separately by group. The solid lines represent the total circular velocity, while the dashed line indicates the contribution from the disk component in each group. In the absence of a bulge, the total rotation curve rises gradually with $R$ and saturates at large radii. It is evident that a galaxy with a less centrally concentrated halo, characterized by a larger scale radius, has a lower circular velocity as a result of the weakened gravitational pull. The total circular velocity measured at $R = 20\kpc$ is provided in Column (8) of \autoref{tbl:model}.

\begin{figure*}[t]
\centering
\epsscale{1.0} \plotone{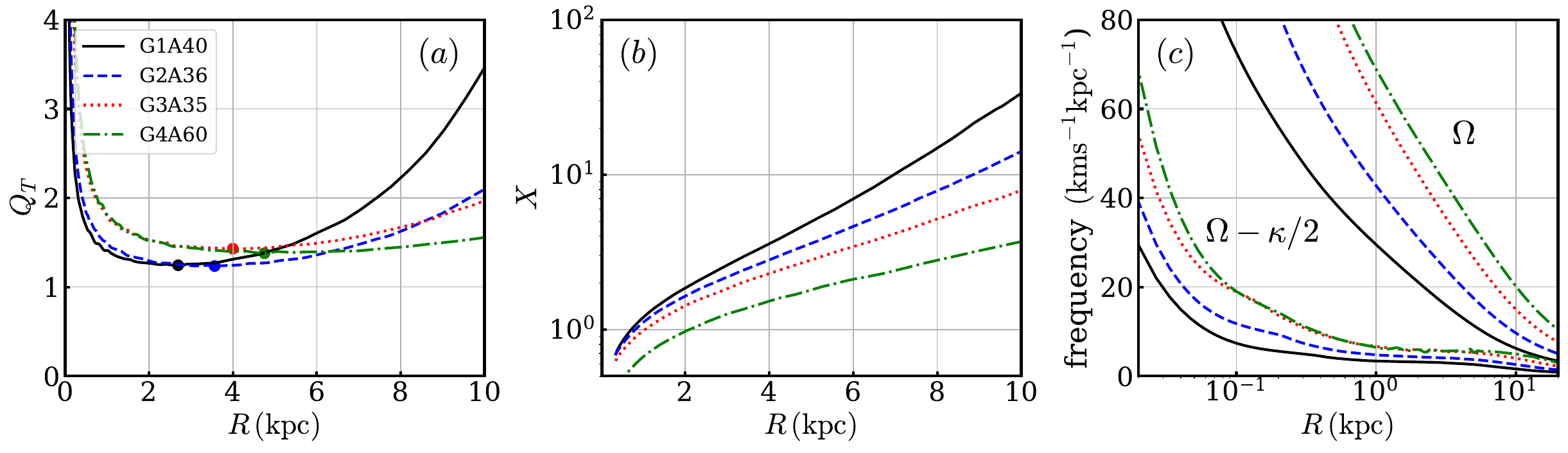}
\caption{Radial distributions of ($a$) the Toomre stability parameter $\QT$, ($b$) the dimensionless azimuthal wavelength $X$, and ($c$) frequencies  $\Omega$ (solid) and  $\Omega-\kappa/2$ (dashed) at $t=0$. In $(a)$, a small dot marks the location of the minimum $\QT$ for each model. 
\label{fig:qtok}}
\end{figure*}

\cref{fig:qtok} plots the radial distributions of $Q_T$, $X$, and frequencies $\Omega$ and $\Omega-\kappa/2$ at $t=0$ for the four fiducial models. Overall, $Q_T$ reaches a minimum at $R\sim (1.1$--$1.4)R_d$, increasing toward smaller radii as $\kappa$ rises and toward larger radii as $\Sigma$ declines. Note that $X$ increases monotonically with radius and is of order unity at $R \sim R_d$. Since $\Omega-\kappa/2$ rises steeply toward smaller radii and diverges at $R=0$, all models exhibit a strong \ac{ILR}, which acts as a barrier to density wave propagation. This suggests that the feedback loop of swing amplification, if present, may be sustained by wave reflections at the \ac{ILR} and/or by nonlinear interactions that regenerate leading waves from trailing ones \citep[e.g.,][]{bnt08}.

\subsection{Numerical Method} 

We construct the initial galaxy models using the GALIC code \citep{yu14}, which solves the collisionless Boltzmann equations by optimizing the velocities of individual particles to achieve the desired equilibrium state. Each model consists of $N_d = 1.0 \times 10^6$ disk particles and $N_h = 2.0 \times 10^7$ halo particles. The galaxy models are evolved for $10\Gyr$ using the public version of the Gadget-4 code \citep{springel21}, following the setup described in \citet{jnk23}. We adopt the Fast Multipole Method with a multipole expansion order of $p = 4$ and employ a hierarchical time-integration scheme to enhance computational efficiency. The gravitational softening lengths are set to $0.05\kpc$ for halo particles and $0.01\kpc$ for disk particles.

\section{Numerical Results} \label{sec:results}

This section presents the outcomes of our galaxy simulations. We first examine swing amplification and the evolution of bars, including their interaction with outer spiral arms. We then analyze bar properties and the development of vertical buckling instability. The conditions for bar formation are addressed in \autoref{sec:condition}.

\subsection{Bar Formation and Evolution}\label{sec:barform}

\begin{figure*}[t]
\centering
\epsscale{1.0} \plotone{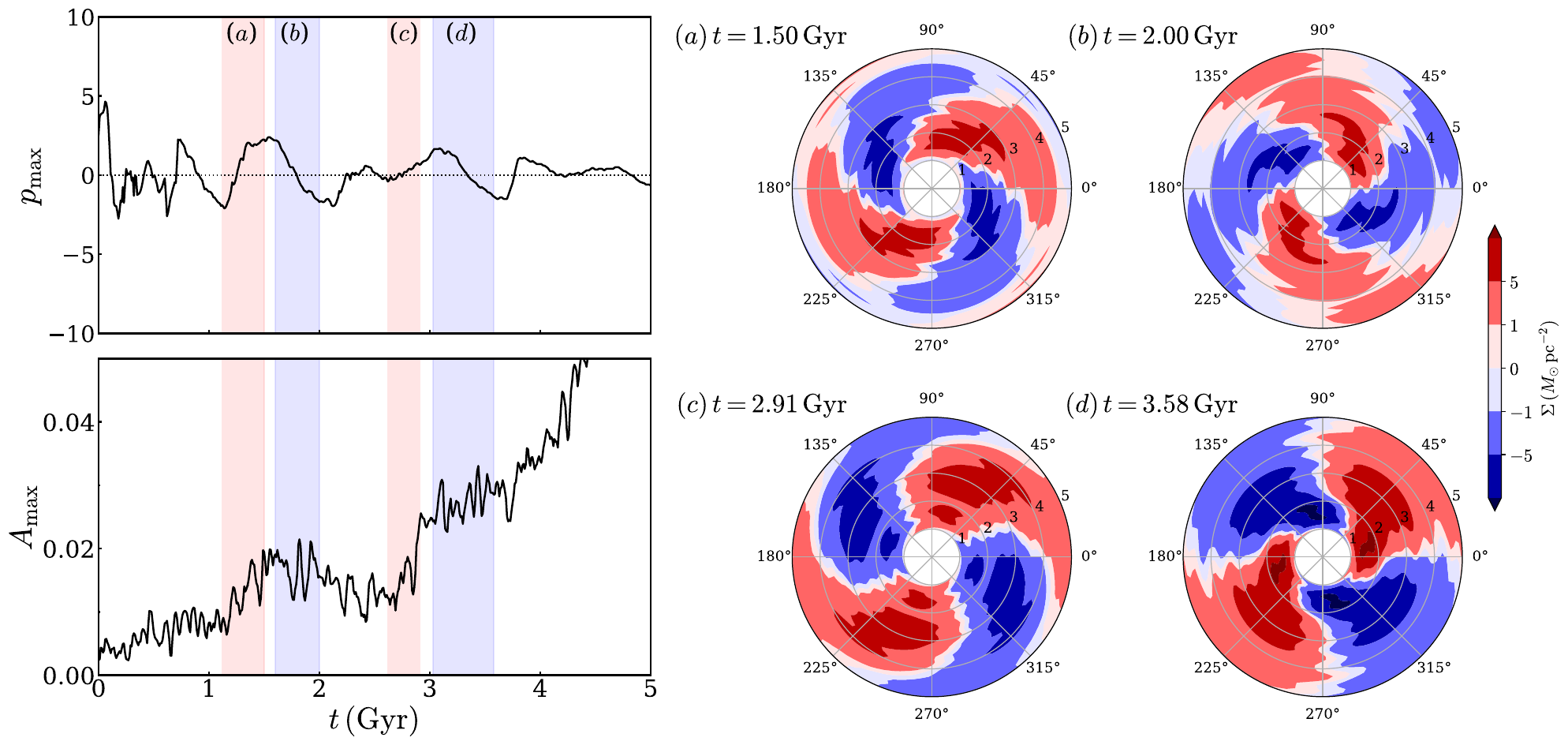}
\caption{Left: Evolution of the maximum wavenumber $p_\text{max}$ that maximizes the Fourier amplitude $|A(2,p)|$, as defined in \cref{eq:Amp}, for the $m = 2$ logarithmic spirals in model \texttt{G3A29},
along with the corresponding maximum amplitude $A_\text{max} = |A(2,p_\text{max})|$. The shaded regions labeled ($a$) and ($c$) indicate the time intervals of swing amplification, while those labeled ($b$) and ($d$) correspond to the feedback stage. Right: Surface density maps of the $m =2$ mode at the final times in the corresponding shaded regions. Rotation is counterclockwise in the disk. 
\label{fig:a2pmax}}
\end{figure*}

In our $N$-body simulations, small-amplitude perturbations inherent in a stellar disk grow as they swing from a leading to a trailing configuration. This swing amplification is most effective in the region where $Q_T$ is low and the disk is rotationally supported but not overly stabilized by random motions. As trailing density waves reflect off the central region, where a strong ILR prevents further inward propagation, they can initiate a feedback loop through nonlinear interactions with other trailing waves \citep[e.g.,][]{bnt08}. This feedback generates leading waves, which are then swing-amplified again as they shear into trailing configurations, further enhancing the perturbations. 

To demonstrate the swing amplification and feedback loop, we define the amplitude of logarithmic spiral density waves with the dimensionless radial wavenumber $p$ and azimuthal mode number $m$ as 
\begin{equation}\label{eq:Amp}
A(m,p) = \frac{1}{N} \sum_{j=1}^N \exp[i(m\phi_j + p \ln R_j)],
\end{equation}
where $N$ is the total number of particles and $(R_j, \phi_j)$ are the radial and azimuthal coordinates of the $j$th particle in the regions with $1\kpc\leq R\leq 5 \kpc$ \citep[e.g.,][]{Sell84,Sell86,Oh08}. Since $p$ is related to the pitch angle $\alpha=\cot^{-1}(p/m)$ for $m$-armed spirals, waves with $p < 0$ correspond to leading spirals, while those with $p > 0$ are trailing.

As an illustrative example of swing amplification and feedback, we select model \texttt{G3A29} which has a prolonged growth phase leading to bar formation. \cref{fig:a2pmax} plots the temporal evolution of $p_\text{max}$, the value of $p$ at which $|A(2,p)|$ is maximized, along with the corresponding amplitude $A_\text{max} = |A(2,p_\text{max})|$ for the $m=2$ mode. The wave amplitude grows during the phase in which $p_\text{max}$ shifts from negative to positive values, while leading waves with negative $p$ are generated through reflections at the \ac{ILR} and nonlinear wave interactions. With $\Qbar=1.7$ and $\Xbar=1.9$, the disk in model \texttt{G3A29} is moderately susceptible to self-gravity, allowing the waves to undergo repeated swing amplification and feedback cycles that ultimately give rise to bar formation.  In highly gravitationally susceptible disks with smaller $\Qbar$, one or two episodes of swing amplification are sufficient to drive the system into a state where the amplified perturbations saturate into a global bisymmetric bar. In contrast, disks with sufficiently large values of $\Qbar$ and/or $\Xbar$ experience only weak swing amplification and feedback, even when recurrent, placing them in a quasi-linear regime characterized by a fluctuating density field and no coherent bar structure.

To quantify the bar strength, we consider an annular region of the disk centered at radius $R$ with a width of $\Delta R=1\kpc$ and calculate the maximum Fourier amplitudes of  the $m = 2$ mode across these annuli as 
\begin{equation}\label{e:barstr}
   \frac{A_2}{A_0} = \max_R \left|
   \frac{\sum_j \mu_j e^{i2\phi_j}}{\Sigma_{j}\mu_j}\right|.
\end{equation}
where $\phi_j$ and $\mu_j$ are the azimuthal angle and mass of the $j$th disk particle in each annulus, respectively, and the summation is performed over all particles within the corresponding annular region \citep{se18,kd18,ghosh23,jnk23,jnk24,ghosh24}.
Following \citet{algorry17}, we classify a structure as a bar when the Fourier amplitude satisfies $A_2/A_0 \geq 0.2$; features with $A_2/A_0 < 0.2$ are interpreted as ovals or spiral arms.
  
\begin{figure*}[t!]
\centering
\epsscale{0.9} \plotone{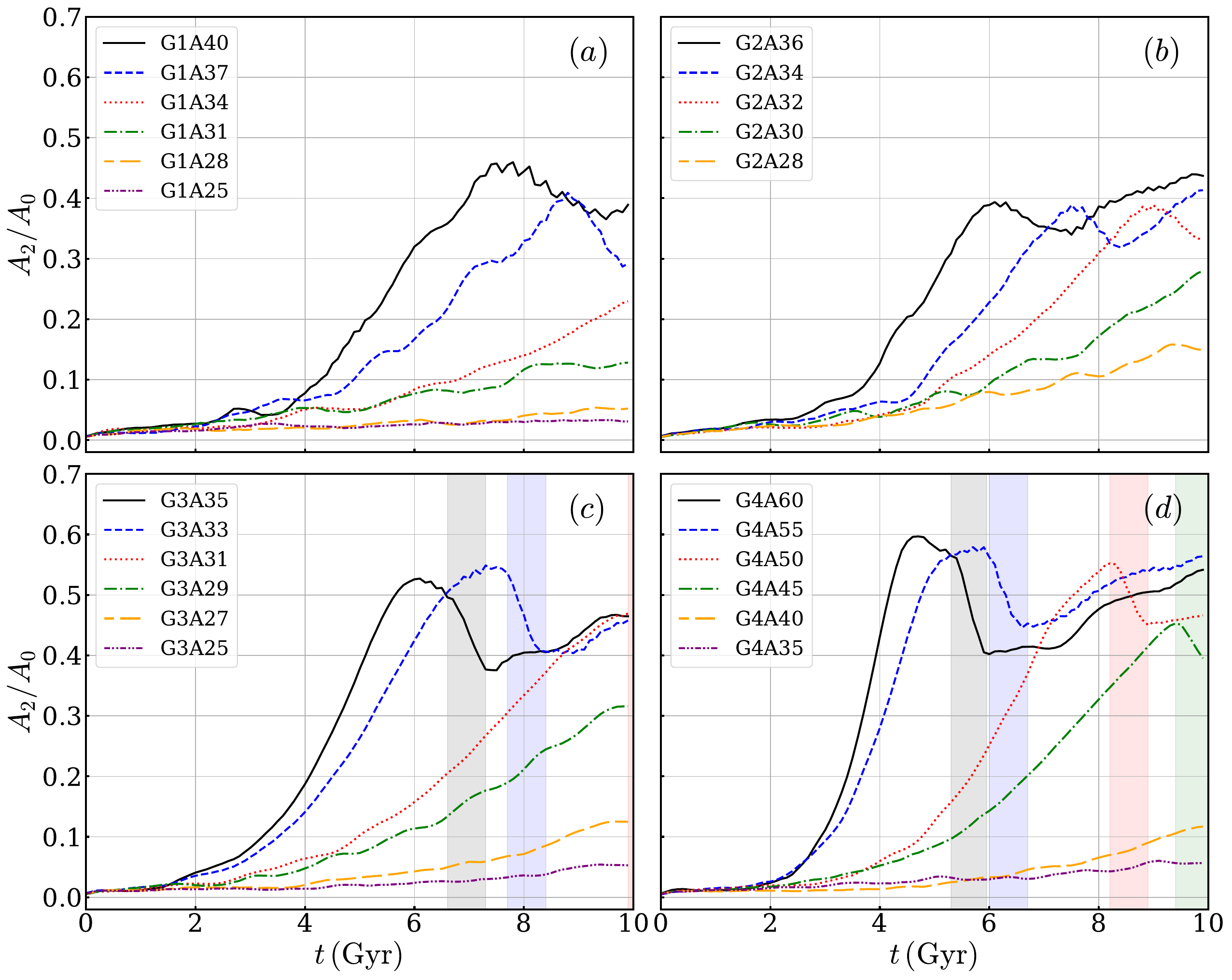}
\caption{Temporal changes of the bar strength $A_2/A_0$ for all models in $(a)$ Group 1, $(b)$ Group 2, $(c)$ Group 3, and $(d)$ Group 4. The shaded vertical bands in ($c$) and ($d$) indicate the periods of buckling instability.
\label{fig:a2a0}}
\end{figure*}

\begin{figure*}[t!]
\centering
\epsscale{0.9} \plotone{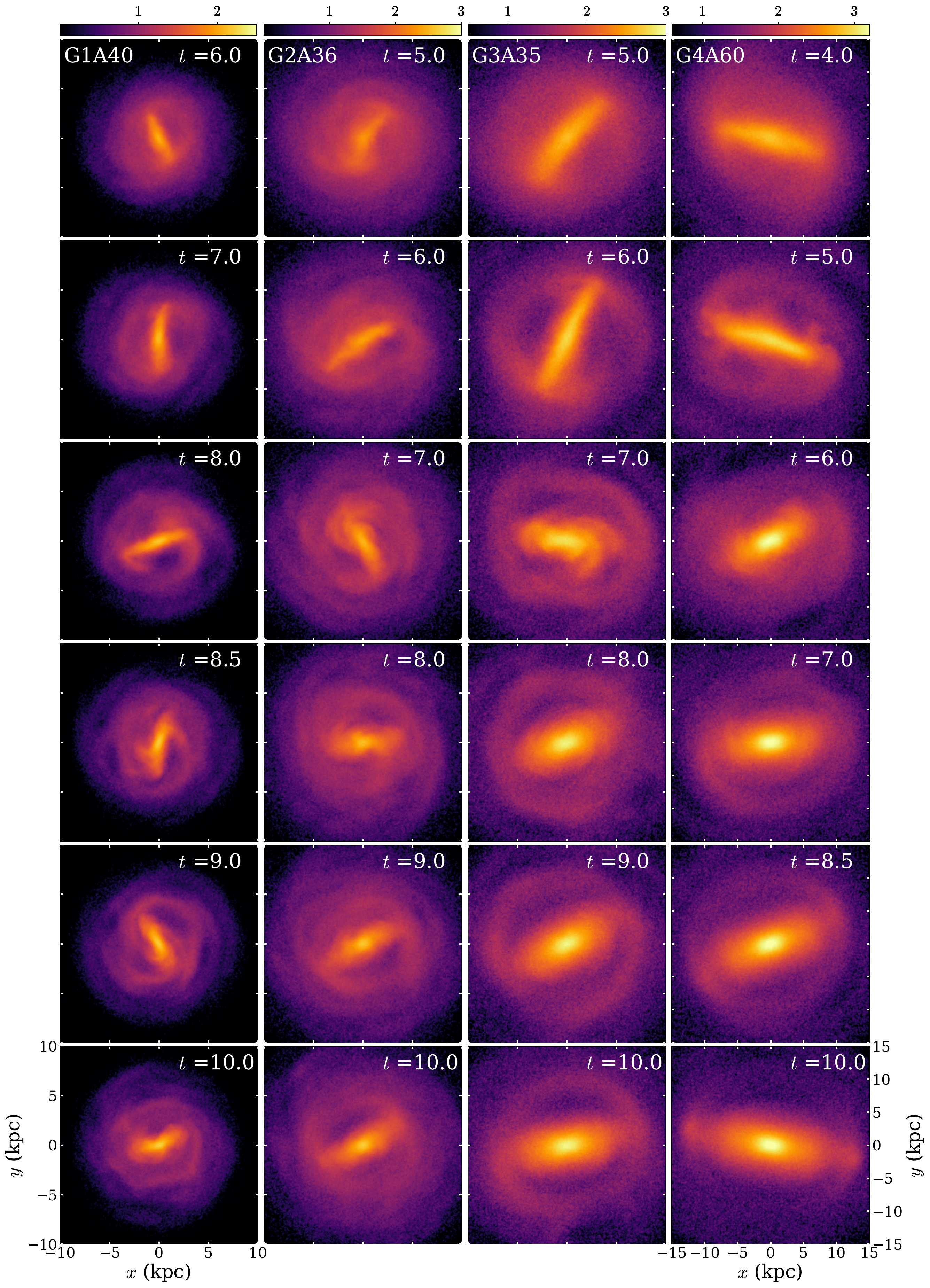}
\caption{Logarithm of the disk surface density $\Sigma$ at selected times for model \texttt{G1A40}, \texttt{G2A36}, \texttt{G3A35} and \texttt{G4A60}, from left to right. The time, $t$, is given in Gyr and the colorbars at the top label $\log \Sigma/({\rm M}_{\odot}\,\text{pc}^{-2})$.
\label{fig:snapshots}}
\end{figure*}

\begin{figure*}[t]
\centering
\epsscale{1.0} \plotone{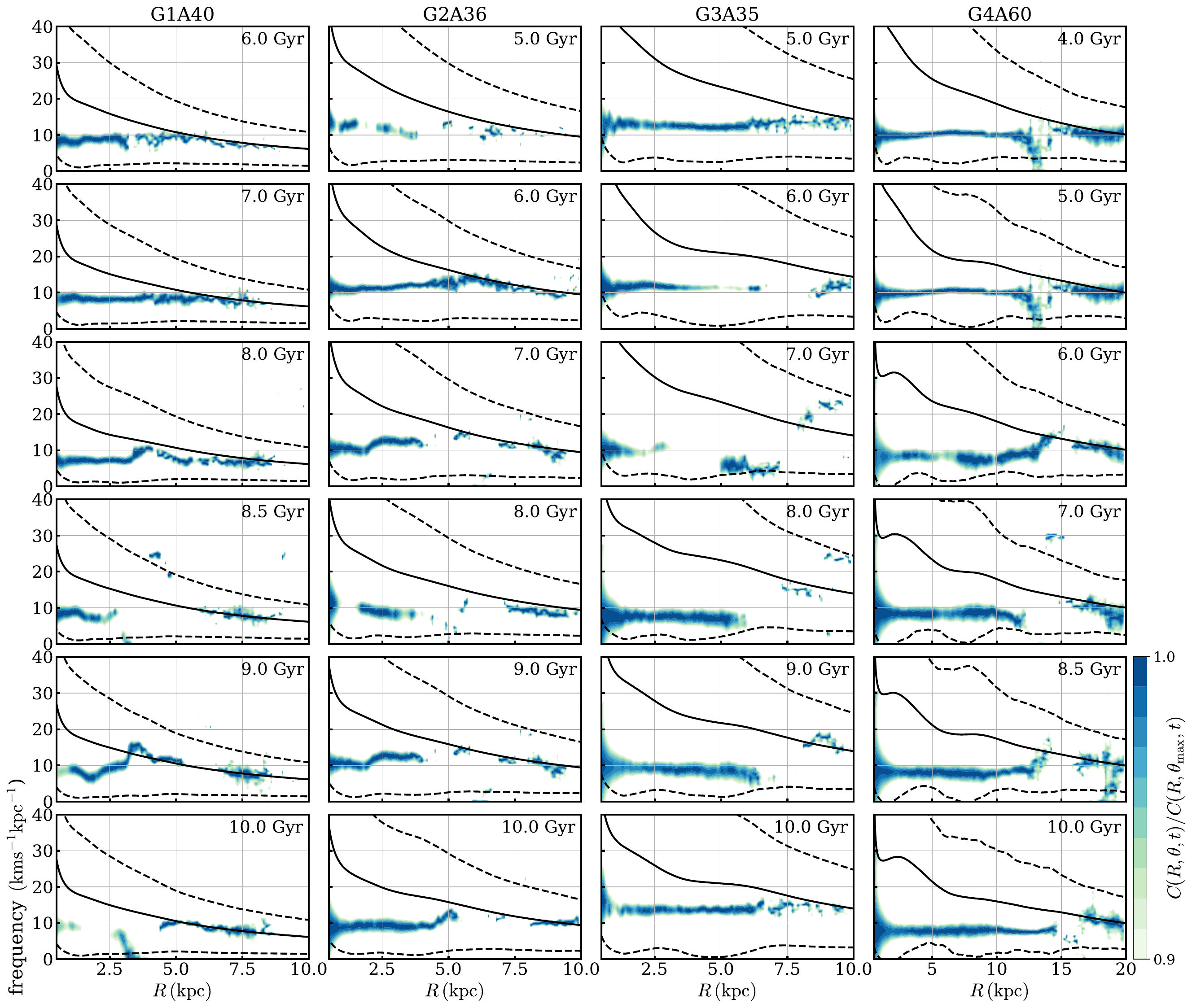}
\caption{Contours of the normalized cross-correlation of the perturbed surface density in the radius-frequency plane for the snapshots shown in \autoref{fig:snapshots}. Various curves draw instantaneous $\Omega$ (solid) and $\Omega\pm\kappa/2$ (dashed) at a given time.
\label{fig:om_omp}}
\end{figure*} 

\autoref{fig:a2a0} plots the temporal evolution of $A_2/A_0$ for all models in our simulation. Overall, a larger halo scale radius $a_h$ leads to earlier and stronger bar formation, and bars tend to be more prominent in galaxies with more massive stellar disks. \cref{fig:snapshots} plots selected snapshots of the disk surface density $\Sigma$  in the fiducial models \texttt{G1A40}, \texttt{G2A36}, \texttt{G3A35}, and \texttt{G4A60}. \cref{fig:om_omp} overlays the frequency curves with the normalized cross-correlation of the perturbed surface densities at two different times, corresponding to the snapshots shown in \cref{fig:snapshots}. Here, the cross-correlation is defined as 
\begin{equation}\label{eq:CC}
\begin{split}
    C(R,\theta,t)&\equiv\frac{1}{\Sigma_0^2}\\
    \times\int^{2\pi}_0& \delta\Sigma (R,\phi,t)\delta \Sigma (R,\phi+\theta,t+\delta t)d\phi ,
    \end{split}
\end{equation}
where $\delta\Sigma \equiv \Sigma -\Sigma_0$ for the initial surface density $\Sigma_0=\Sigma(t=0)$ and $\delta t=0.1 \Gyr$
\citep[e.g.,][]{Oh08,oh15,seo19}.
The instantaneous pattern speed of a bar or spiral arms at a given radius is then determined by $\Omega_b(R,t)=\theta_\text{max}/\delta t$, where $\theta_\text{max}$ denotes the phase angle at which $C(R,\theta,t)$ is maximized. Strictly speaking, a genuine pattern is characterized by a constant $\Omega_b$, with the radii satisfying $\Omega_b=\Omega$ and $\Omega_b=\Omega\pm\kappa/2$ corresponding to the corotation and Lindblad resonances, respectively. 

In Group 1, model \texttt{G1A25} shows no significant increase in $A_2/A_0$ up to $10\Gyr$, indicating no bar formation in this model. Within this group, bar formation occurs only in models with $a_h \geq 34 \kpc$, where bars grow slowly and remain relatively weak and short ($\sim2$--$3\kpc)$. Beyond the bar region, the disk hosts relatively strong spiral arms that nearly corotate with the bar at early times, as indicated by $\Omega_b\sim \Omega$ at $R\sim(5$--$8)\kpc$ and $t\sim6$--$7\Gyr$. At later times, however, the spirals evolve differently, interacting with the bar episodically in either a constructive or destructive manner. In model \texttt{G1A40}, for example, the spiral arms align their phases with that of the bar, allowing it to reach a maximum $A_2/A_0$ at $t \sim 8 \Gyr$ \citep[e.g.,][]{jnk23}. Subsequently, the arms interfere with the bar destructively, 
reducing the extent of the regions with constant $\Omega_b$ and thereby making the bar shorter and weaker. \cref{fig:om_omp} shows that at $t = 8\Gyr$, model \texttt{G1A40} has a bar pattern speed of $\Omega_b \sim 8\freq$, while the outer spiral arms exhibit a pattern speed close to $\Omega$, suggesting that they are material arms. Progressive interference from the spiral arms leads to the dissolution of the bar as a coherent structure, with no distinct bar pattern discernible at $t \gtrsim 8.5\Gyr$. We therefore conclude that the bar in model \texttt{G1A40} is disrupted by the spiral arms around this time.

\begin{figure*}[t]
\centering
\epsscale{0.9} \plotone{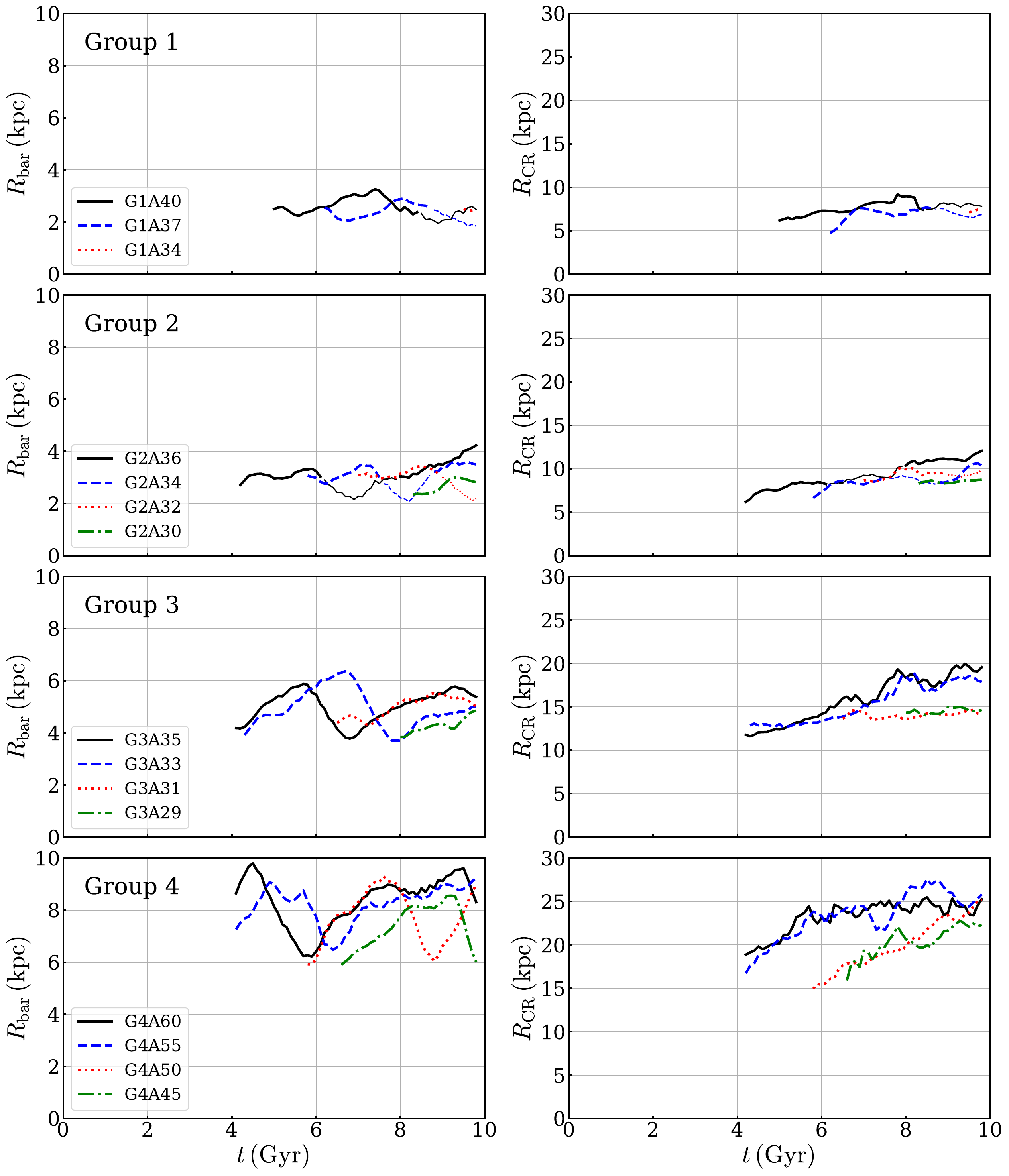}
\caption{Temporal evolution of the bar length $R_\text{bar}$ (left) and $R_\text{CR}$ (right) for the bar-forming models in Group 1, Group 2, Group 3, and Group 4 from top to bottom. The thin portions of the curves for models \texttt{G1A37}, \texttt{G1A40}, \texttt{G2A32}, \texttt{G2A34}, and \texttt{G2A36} indicate phases during which the bars are disrupted by outer spiral arms. 
\label{fig:lbar}}
\end{figure*}

\begin{figure*}[t]
\centering
\epsscale{0.9} \plotone{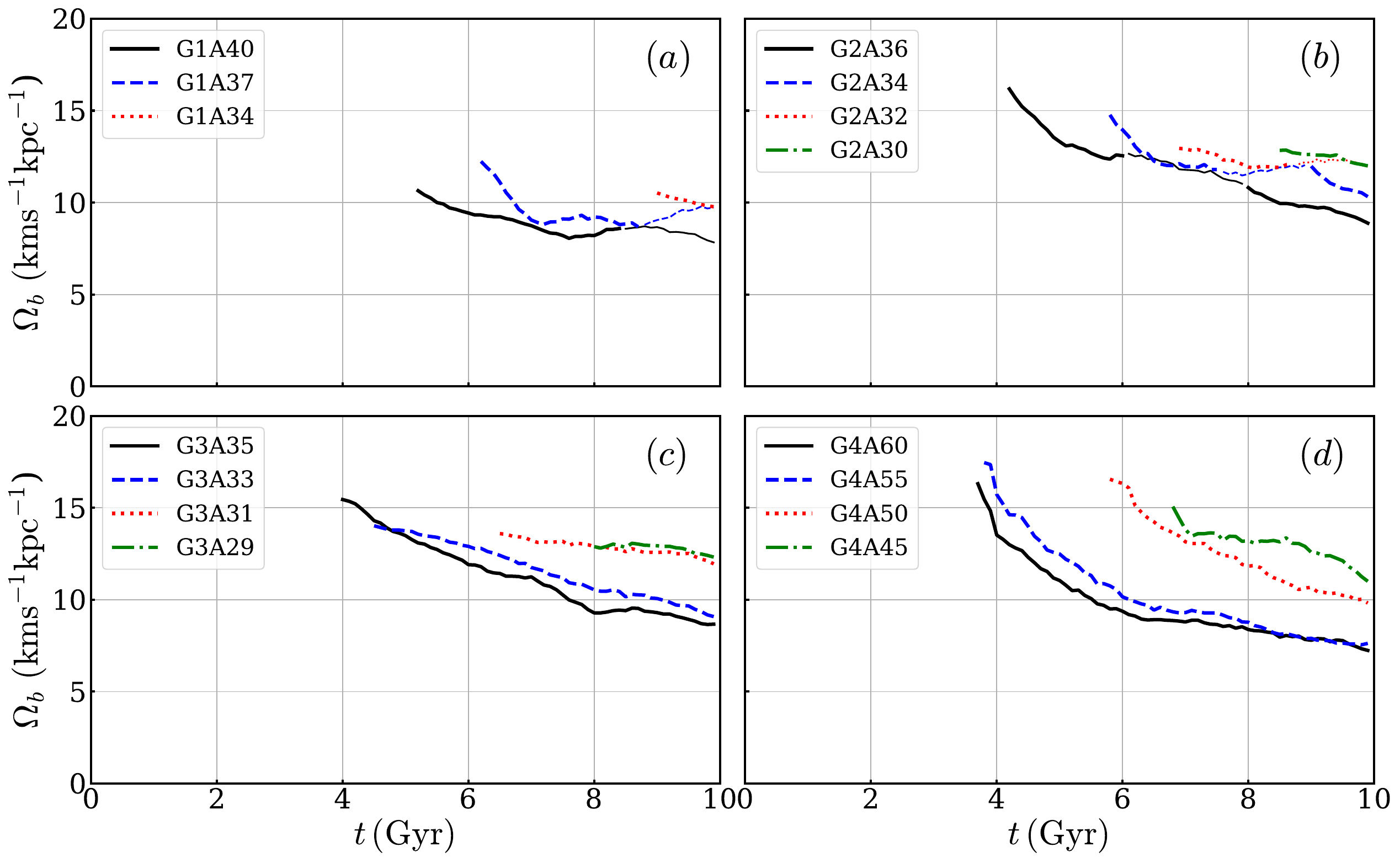}
\caption{Temporal evolution of the bar pattern speed $\Omega_b$ for the bar-forming models
in $(a)$ Group 1, $(b)$ Group 2, $(c)$ Group 3, and $(d)$ Group 4. The thin portions of the curves for models \texttt{G1A37}, \texttt{G1A40}, \texttt{G2A32}, \texttt{G2A34}, and \texttt{G2A36} indicate phases during which the bars are disrupted by outer spiral arms. 
\label{fig:omega}}
\end{figure*}

A similar situation occurs in Group 2, although the bars form earlier and attain greater length and strength compared to those in Group 1. All models in this group develop a bar, except for model \texttt{G2A28}, which shows merely a faint oval distortion confined to the central region of the disk. In model \texttt{G2A36}, $A_2/A_0$ reaches its maximum at $t \sim 6\Gyr$, when the phase angle of the bar aligns with that of the outer spiral arms. The ensuing destructive interactions between the bar and spiral arms not only shorten the bar but also modify the radial profile of $\Omega_b$. Since the outer spiral arms are relatively weak in Group 2, however, the destructive interactions are temporary: the bar retains its identity at small radii and begins to grow again at $t \gtrsim 9\Gyr$, recovering a constancy of $\Omega_b$ up to $R\sim4\kpc$ at $t=10\Gyr$. The evolution of the other bar-forming models is similar, with bar formation occurring later in models with smaller $a_h$.

In Group 3, all models except models \texttt{G3A27} and \texttt{G3A25} develop a bar, with formation proceeding more rapidly in those with larger $a_h$. The bars in this group form earlier and are longer ($\sim 4$--$6\kpc$) compared to those in Groups 1 and 2.  This leads to two notable differences. First, the spiral arms outside the bar are weak, leading to only limited destructive interference. For instance, \autoref{fig:om_omp} shows that the bar in model \texttt{G3A35} temporarily shortens near  $t = 7\Gyr$ due to interference from the spiral arms, exhibiting a pattern speed of $\Omega_b \sim 12\freq$ confined to the region $R \lesssim 4\kpc$, before regrowing rapidly thereafter. Second, the strong bars in this group undergo buckling instability, becoming weaker and shorter, a process discussed further in \autoref{subsec:buckling}.

\begin{figure*}[t]
\centering
\epsscale{0.9} \plotone{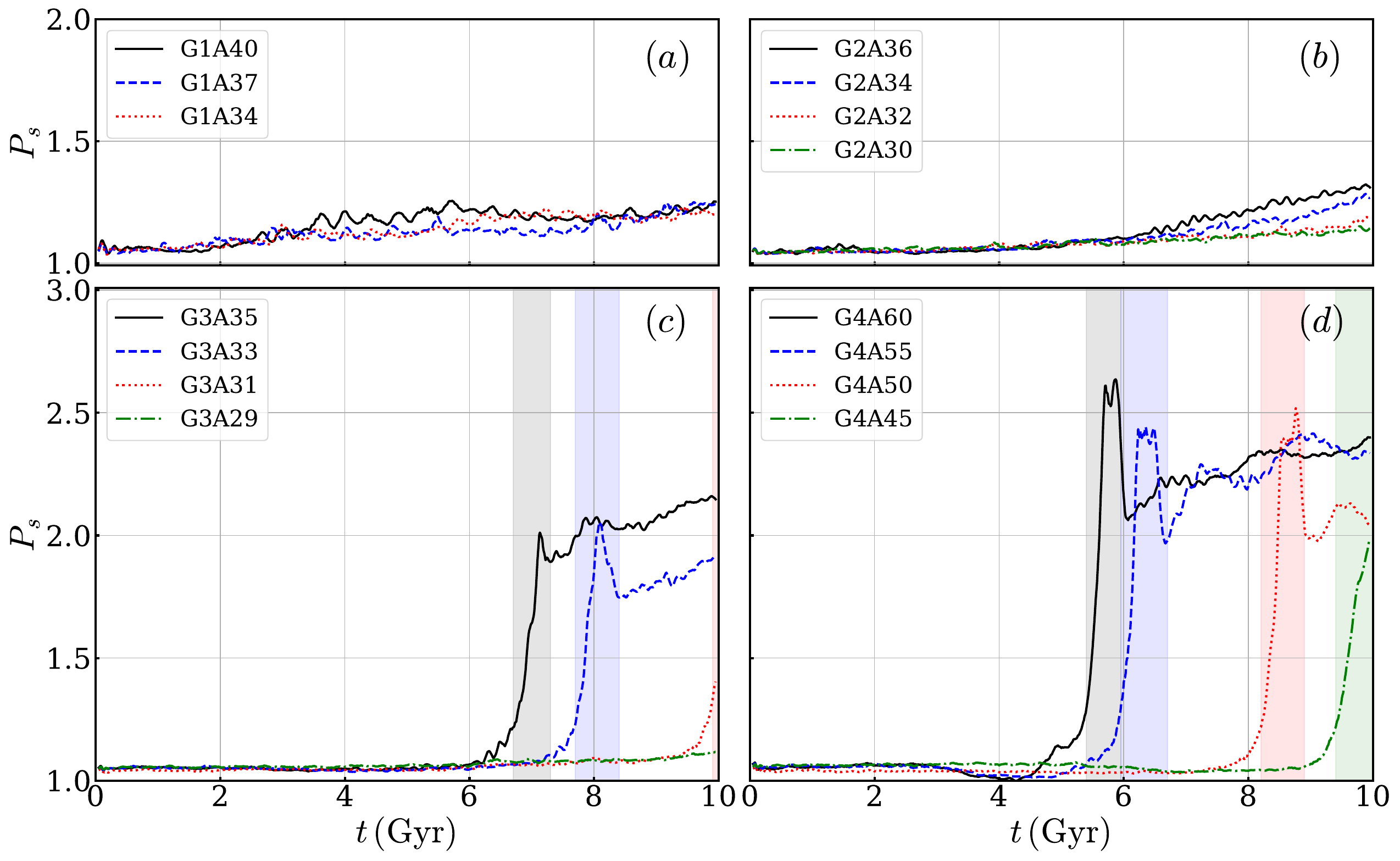}
\caption{Temporal evolution of the BPS strength $P_s$ for all bar-forming models in ($a$) Group 1, ($b$) Group 2, ($c$) Group 3, and ($d$) Group 4. In ($c$) and ($d$), the shaded vertical bands indicate the periods of buckling instability.
\label{fig:ps}}
\end{figure*}

\begin{figure*}[t]
\centering
\epsscale{0.9} \plotone{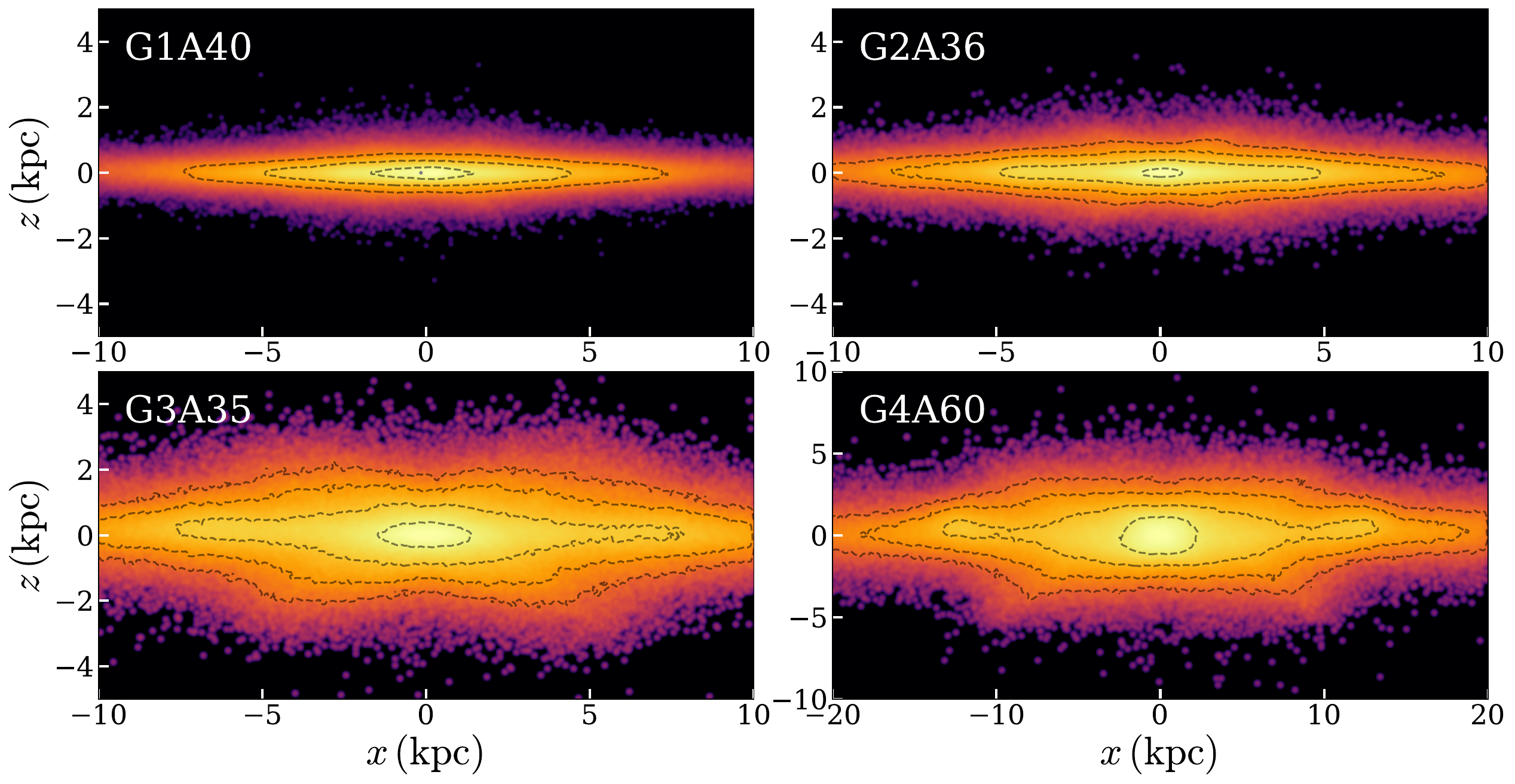}
\caption{Contours of the logarithm of the projected disk densities at the end of the runs ($t = 10 \Gyr$) for the fiducial models. The $x$- and $z$-axes correspond to the semimajor axis of the bar and the vertical direction, respectively. Dotted contours represent projected densities $\int \rho_d \,dy = 10^{9.0}$, $10^{8.5}$, $10^{8.0}$, and $10^{7.5}\Msun\;\kpc^{-2}$, from innermost to outermost.
\label{fig:xz}}
\end{figure*}

\begin{figure*}[t]
\centering
\epsscale{0.9} \plotone{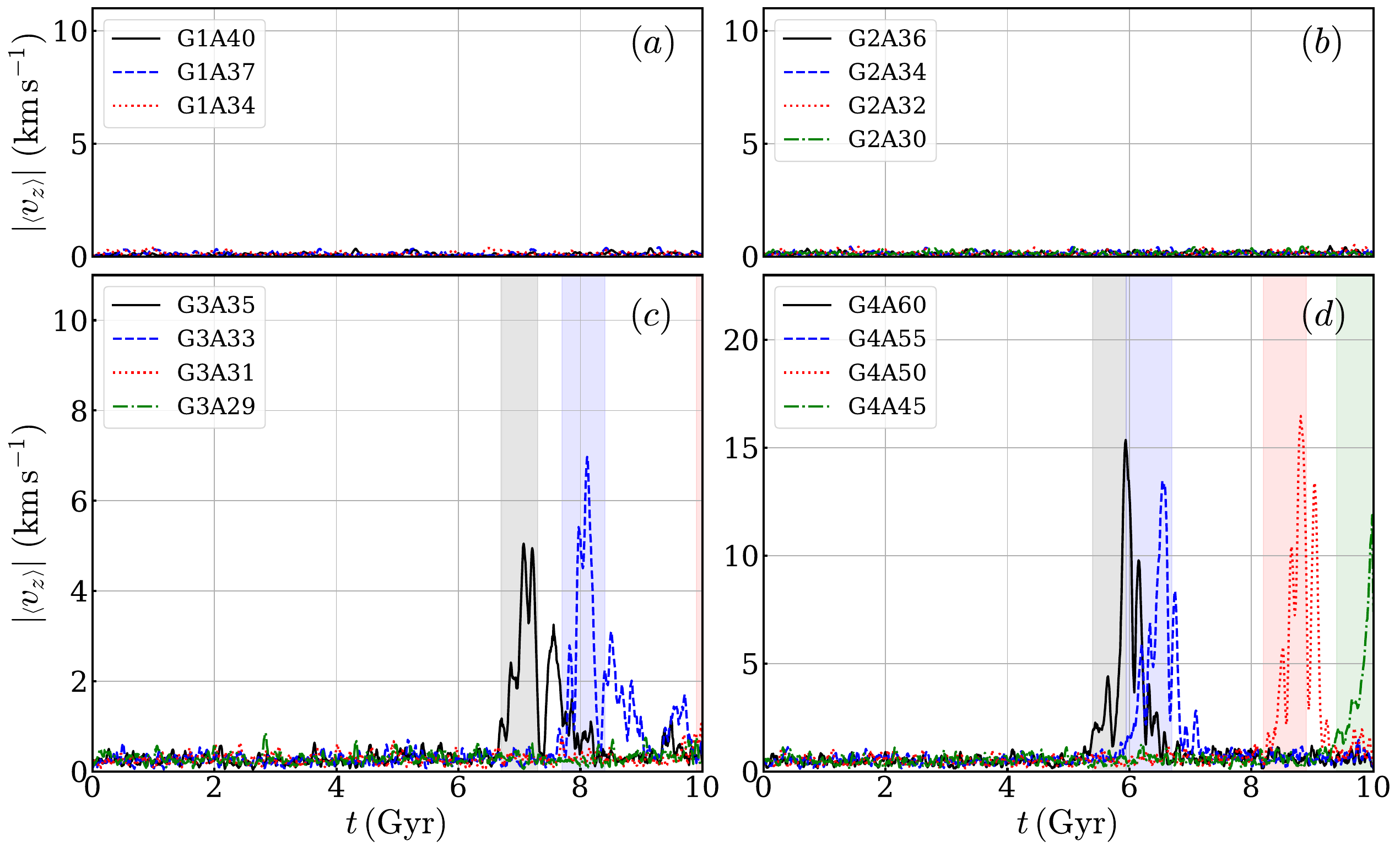}
\caption{Temporal evolution of the mean vertical velocity, $|\langle v_z \rangle|$, measured in an annulus centered at $R = 3\kpc$ for all bar-forming models in ($a$) Group 1, ($b$) Group 2, ($c$) Group 3, and ($d$) Group 4. In ($c$) and ($d$), the shaded vertical bands indicate the duration of the buckling instability. 
\label{fig:vz}}
\end{figure*}

\begin{figure*}[t]
\centering
\epsscale{0.9} \plotone{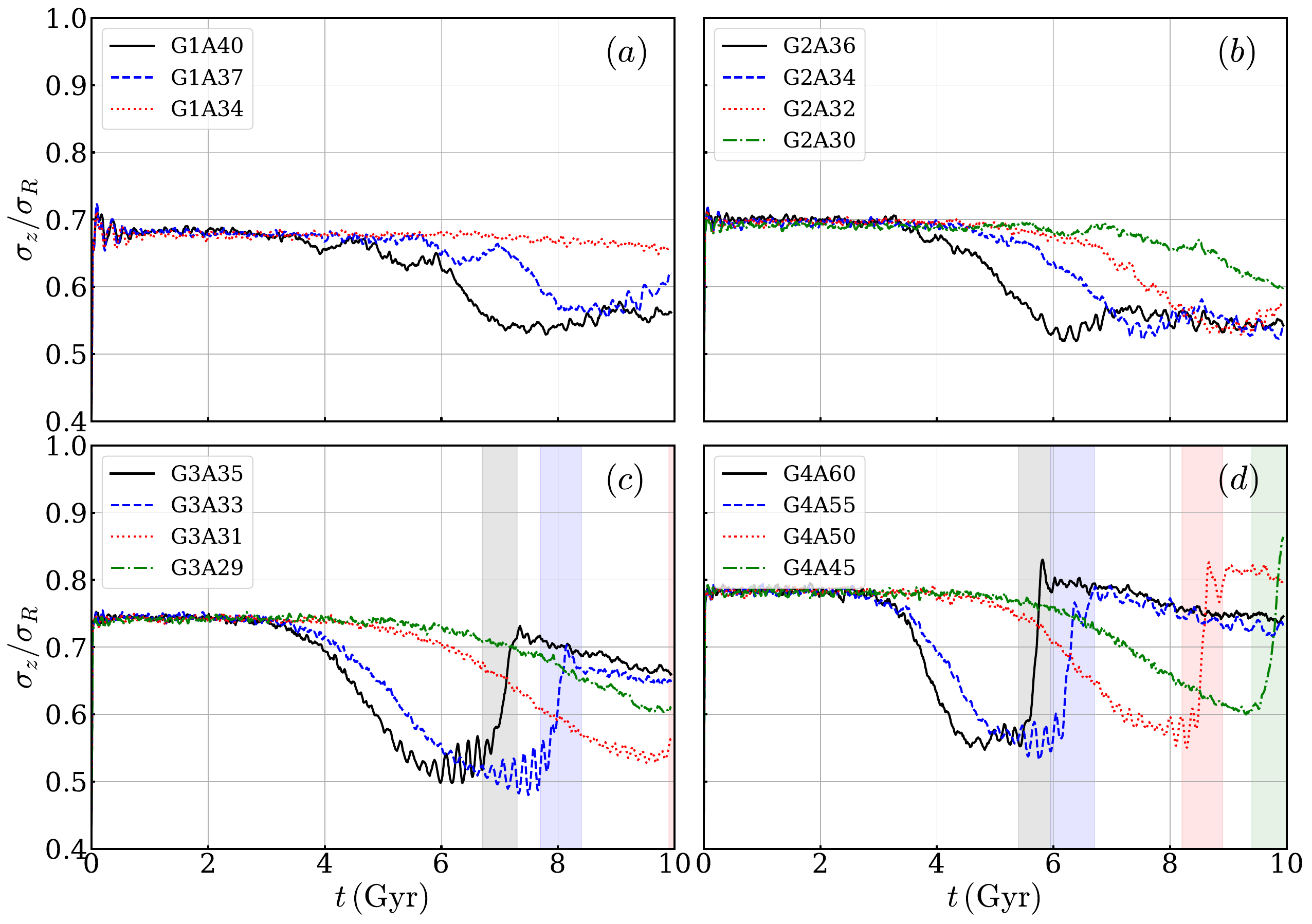}
\caption{Temporal evolution of the ratio $\sigma_z/\sigma_R$ of the vertical to radial velocity dispersions of the disk particles at $R=2\kpc$ for all bar-forming models in ($a$) Group 1, ($b$) Group 2, ($c$) Group 3, and ($d$) Group 4. 
In ($c$) and ($d$), the shaded vertical bands indicate the duration of the buckling instability.
\label{fig:sig}}
\end{figure*}

In Group 4, the threshold value of the halo scale radius for bar formation increases to $a_h \sim 40\kpc$.  Owing to their shorter dynamical times, models in this group develop bars more rapidly than those in the other groups. The bars in this group also grow strong and extended, dominating the regions with $R \lesssim 6$--$10\kpc$. Since the outer spirals are relatively weak, their impact on the dynamical evolution of the bars is minimal. As a result, the bars preserve their identity as coherent patterns throughout the simulations, as evidenced by the broad radial range of constant $\Omega_b$ shown in \cref{fig:om_omp}. As in Group 3, the models in Group 4 exhibit a sudden drop in $A_2/A_0$ due to buckling instability, after which the bars regrow in size and strength, most likely through the trapping of stars into bar-supporting orbits \citep[e.g.,][]{mar06}.
\subsection{Bar Length and Pattern Speed} \label{subsec:omega}

To determine the bar length at a given time, we calculate the bar position angle
\begin{equation}\label{e:posang}
 \psi(R)\equiv \frac{1}{2} \tan^{-1}\left[\frac{\sum_j \mu_j \cos(2\phi_j)}{\sum_j \mu_j \sin(2\phi_j)}\right],
\end{equation} 
as a function of $R$. The bar length, $R_\text{bar}$, is then defined as the radius within which the variation in $\psi(R)$ remains less than 0.1 radians \citep{athmis02,scannapieco12,jnk23}. \autoref{fig:lbar} presents the temporal evolution of $R_\text{bar}$ and the corotation radius, $R_\text{CR}$, in the models that forms a bar. Thin portions of the curves for models \texttt{G1A37}, \texttt{G1A40}, \texttt{G2A32}, \texttt{G2A34}, and \texttt{G2A36} mark intervals of bar disruption caused by outer spirals. In general, all bars in our models lie well inside $R_\text{CR}$. They tend to be longer in galaxies with more massive stellar disks, with $R_\text{bar} \sim 2$--$3\kpc$ in Group 1, $\sim2$--$4\kpc$ in Group 2, $\sim4$--$6\kpc$ in Group 3, and $\sim 6$--$10\kpc$ in Group 4. Within each group, models with more compact halos (i.e., smaller $a_h$) initially form shorter bars. However, dynamical processes such as spiral interactions, buckling instability, and bar regrowth render the late-time dependence of $R_\text{bar}$ on $a_h$ more complex. Overall, the temporal evolution of the bar length qualitatively mirrors that of the bar strength, as shown in \autoref{fig:a2a0}. Notably, in Groups 3 and 4, which have massive stellar disks, bars undergo significant structural changes in length due to buckling instability, which is not observed in the low-mass disk models.

As mentioned above, the cross-correlation of the perturbed surface densities, defined in \cref{eq:CC}, provides a useful means of determining the bar pattern speed $\Omega_b(R,t)$. \autoref{fig:omega} plots the temporal evolution of $\Omega_b$, measured at $R=2\kpc$, for all bar-forming models. The initial pattern speed at the time of bar formation tends to be higher in models with more massive disks. In most cases, the gradual decline of $\Omega_b$ over time results from angular momentum transfer to the surrounding live halo, with the amount of slowdown being greater in models with more massive disks. In some low-mass models, such as \texttt{G1A37} and \texttt{G2A34}, interactions with outer spiral arms can temporarily enhance $\Omega_b$; however, this effect is absent in the more massive galaxies of Groups 3 and 4.

\subsection{Disk Thickening and Buckling Instability}\label{subsec:buckling}

In our models, bars thicken vertically over time and evolve into boxy/peanut-shaped (BPS) bulges when sufficiently strong. The strength of the BPS structure is typically quantified by
\begin{equation}\label{eq:Ps}
P_s \equiv \max_R \left( \frac{\widetilde{|z|}}{\widetilde{|z_0|}}\right),
\end{equation}
where the tilde denotes the median, and $z_0$ is the initial vertical height of disk particles within $R\leq 10\kpc$ \citep{ian15,frag17,seo19,jnk23,jnk24}. \autoref{fig:ps} plots the temporal variations of $P_s$ for all bar-forming models.
\autoref{fig:xz} plots the edge-on projections of the stellar density at the end of the runs for the fiducial models, with the bar semimajor axes aligned along the horizontal axis. Bars in Groups 1 and 2 exhibit only mild vertical thickening throughout the entire evolution, reflecting their relatively weak amplitudes. In contrast, the strong bars formed in Groups 3 and 4 undergo abrupt vertical thickening at certain epochs, a consequence of the buckling instability that temporarily disrupts vertical symmetry and redistributes stellar orbits.

The onset of buckling instability can be effectively quantified by monitoring the evolution of $|\langle v_z \rangle|$, where the angle brackets denote an angular average over a radial annulus of width $\Delta R = 0.1\kpc$ centered at $R = 3.0\kpc$ \citep{kwak17,kwak19}.   \autoref{fig:vz} plots the temporal evolution of $|\langle v_z \rangle|$ for all bar-forming models. Evidently, all bars in Groups 1 and 2 maintain $|\langle v_z \rangle| \simeq 0$ throughout the entire evolution, indicating that buckling instability does not occur in these models. Consequently, the disks remain nearly symmetric about the $z=0$ plane. However, models \texttt{G3A35}, \texttt{G3A33}, and \texttt{G3A31} in Group 3 undergo buckling episodes around $t \sim 6.7$--$7.3\Gyr$, $\sim 7.7$--$8.4\Gyr$, and $\gtrsim 9.9\Gyr$, respectively.
Models in Group 4 experience stronger buckling instability at $t\sim 5.4$--$6.0\Gyr$, $\sim5.9$--$6.7\Gyr$, $\sim8.2$--$8.9\Gyr$, and $\gtrsim 9.4\Gyr$, for models \texttt{G4A60}, \texttt{G4A55}, \texttt{G4A50}, and \texttt{G4A45}, respectively. These episodes, marked by the shaded vertical bands in \autoref{fig:vz}, coincide with the phases of rapid decline in $A_2/A_0$ (see \autoref{fig:a2a0}) and the corresponding sharp increase in $P_s$ (see \autoref{fig:ps}). \autoref{fig:xz} shows that the disks in these models become asymmetric with respect to the $z=0$ plane after experiencing the buckling instability. 

As a physical cause of buckling instability, the firehose instability has often been invoked \citep{toomre66,bnt08}. In this scenario, bending perturbations become unstable when the centrifugal force over vertical corrugations exceeds the restoring gravitational force, which occurs when $\sigma_z/\sigma_R$ drops below a critical value \citep[e.g.,][]{toomre66, ara87}. This critical threshold is $\sim 0.3$ for infinitesimally thin, non-rotating disks and increases to $\sim 0.5$--$0.7$ for realistic, Milky Way–sized galactic disks \citep{mar06,kwak17,seo19,jnk23,jnk24}. It remains unclear whether a similar condition is required for buckling instability in disks of differing mass.

\autoref{fig:sig} plots the temporal evolution of $\sigma_z/\sigma_R$ evaluated at $R = 2\kpc$ for all bar-forming models. The growth of a bar is accompanied by an increase in $\sigma_R$, primarily driven by the prevalence of $x_1$ orbits aligned with its major axis. At later stages, once the bar is fully developed, vertical perturbations originating from both the bar and outer spiral arms can enhance stellar motions perpendicular to the disk, thereby increasing $\sigma_z$ \citep[e.g.,][]{qui14}. In Groups 3 and 4, which contain massive disks, bars become unstable to buckling when $\sigma_z/\sigma_R \lesssim 0.6$, in agreement with the thresholds found for Milky Way–sized galaxies. However, the same condition does not apply to galaxies in Groups 1 and 2: bars in these groups remain stable to buckling even when $\sigma_z/\sigma_R < 0.55$. In low-mass models, interference from outer spiral arms may contribute to the suppression of buckling instability. Alternatively, the bars in these systems grow slowly and remain weak, possibly resulting in a lower critical threshold for instability. These results suggest that a low $\sigma_z/\sigma_R$ may be a necessary but not sufficient condition for the onset of buckling.

\section{Criteria for Bar Formation }\label{sec:condition}

In this section, we apply a local analytic theory of swing amplification to compute the amplification factor as a function of $Q_T$ and $X$. The analytic predictions are then compared with numerical simulations, based on which we propose a criterion for bar formation.

\subsection{Amplification Factor}\label{sec:amp}

\begin{figure}[t]
\centering
\epsscale{1} \plotone{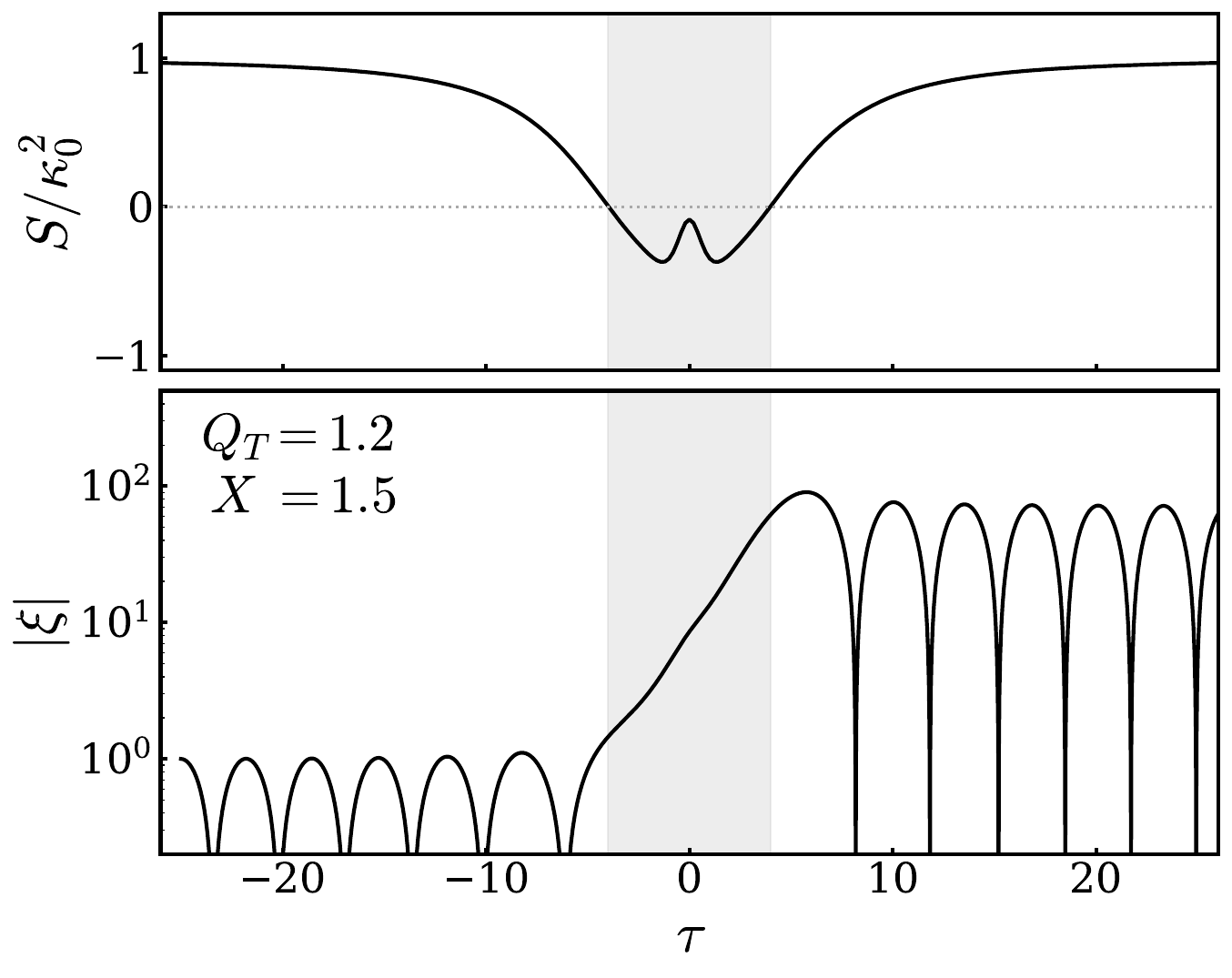}
\caption{Variations of $S / \kappa_0^2$ and $|\xi|$ as functions of dimensionless time $\tau = t \kappa_0$ for $\QT = 1.2$ and $X = 1.5$. The shaded regions denote the period of most significant growth, during which $S<0$. 
\label{fig:snxi}}
\end{figure}

Swing amplification is a phenomenon in shearing, self-gravitating disks whereby leading spiral perturbations are transformed into trailing ones and amplified through the combined effects of differential rotation, epicyclic shaking, and self-gravity \citep{toomre81}. Consider a razor-thin disk characterized by surface density $\Sigma_0$, angular rotational frequency $\Omega_0$, dimensionless shear rate $q\equiv - d\ln \Omega_0/d\ln R$, and epicycle frequency $\kappa_0=(4-2q)^{1/2}\Omega_0$. The disk is assumed to be infinite in extent along both the $x$- and $y$-directions. For a local sinusoidal wavelet with wavenumber $k_y$ along the $y$-direction, the background shear causes the radial wavenumber to increase linearly with time as 
\begin{equation}
k_x(t)=k_x(0)+q\Omega_0 k_y t.
\end{equation}
The total wavenumber is given by $k(t)=(k_x^2+k_y^2)^{1/2}$.
\citet{toomre81} showed that the perturbed displacement $\xi$ of a particle in the wavelet, measured normal to the wavefront and governing the perturbed surface density, obeys the differential equation 
\begin{equation}\label{eq:xi}
    \ddot{\xi}+S(t)\xi = 0,
\end{equation}
where
\begin{equation}\label{eq:s}
    S(t) = \kappa'^2(t)-2\pi G\Sigma_0k(t)\mathcal{F}(\nu,x)
\end{equation}
with the modified epicycles frequency
\begin{equation}\label{eq:kappa}    \kappa'^2=\kappa_0^2+3q^2\Omega_0^2\frac{k_y^4}{k^4} -4q\Omega_0^2\frac{k_y^2}{k^2},
\end{equation}
\citep[see also][]{GLB65,michi16}. In \cref{eq:s}, $\mathcal{F}(\nu,x)$ is the reduction factor that accounts for the stabilizing effect of random stellar motions, defined as 
\begin{equation}\label{eq:F}
    \mathcal{F}(\nu,x)=2(1-\nu^2)\frac{e^{-x}}{x}\sum_{n=1}^{\infty}\frac{I_n(x)}{1-\nu^2/n^2},
\end{equation}
where $\nu^2 \equiv S(t)/\kappa_0^2$ and $x \equiv  {\sigma_R^2 k^2}/{\kappa_0^2}$.

We define the dimensionless time as $\tau\equiv t\kappa_0$, set $k_x(0)=0$ and $q=1$, and express the wavenumber $k_y=m/R$ in terms of the (inverse of the) parameter $X$ in \cref{eq:X}. Under these conditions, \cref{eq:xi} is fully characterized by two dimensionless parameters, $Q_T$ and $X$. By fixing $\xi = \xi_\text{0}=1$ and $\dot{\xi}=0$ at $\tau=-30$ as the initial conditions, we integrate \cref{eq:xi} from $\tau=-30$ to $30$.  \cref{fig:snxi} plots the dependence of $S/\kappa_0^2$ on $\tau$ and the time evolution of $|\xi|$ for a disk with $\QT =1.2$ and $X = 1.5$. For $|\tau| \gg 1$, $S/\kappa_0^2 \sim 1$, indicating that the wavelet undergoes epicyclic oscillations with a period of $2\pi/\kappa_0$. As $|\tau| \to 0$, $S/\kappa_0^2$ decreases and can even become negative, leading to a growth of $|\xi|$ by about two orders of magnitude. Physically, this arises because the wavelet rotates from a leading to a trailing configuration, in the same direction as the epicyclic motions. This alignment causes particles in the wave crests to linger longer in regions of enhanced density, thereby increasing their exposure to self-gravity. The resulting extended interaction significantly amplifies the wave amplitude \citep{toomre81}.

\begin{figure}[t]
\centering
\epsscale{1.0} \plotone{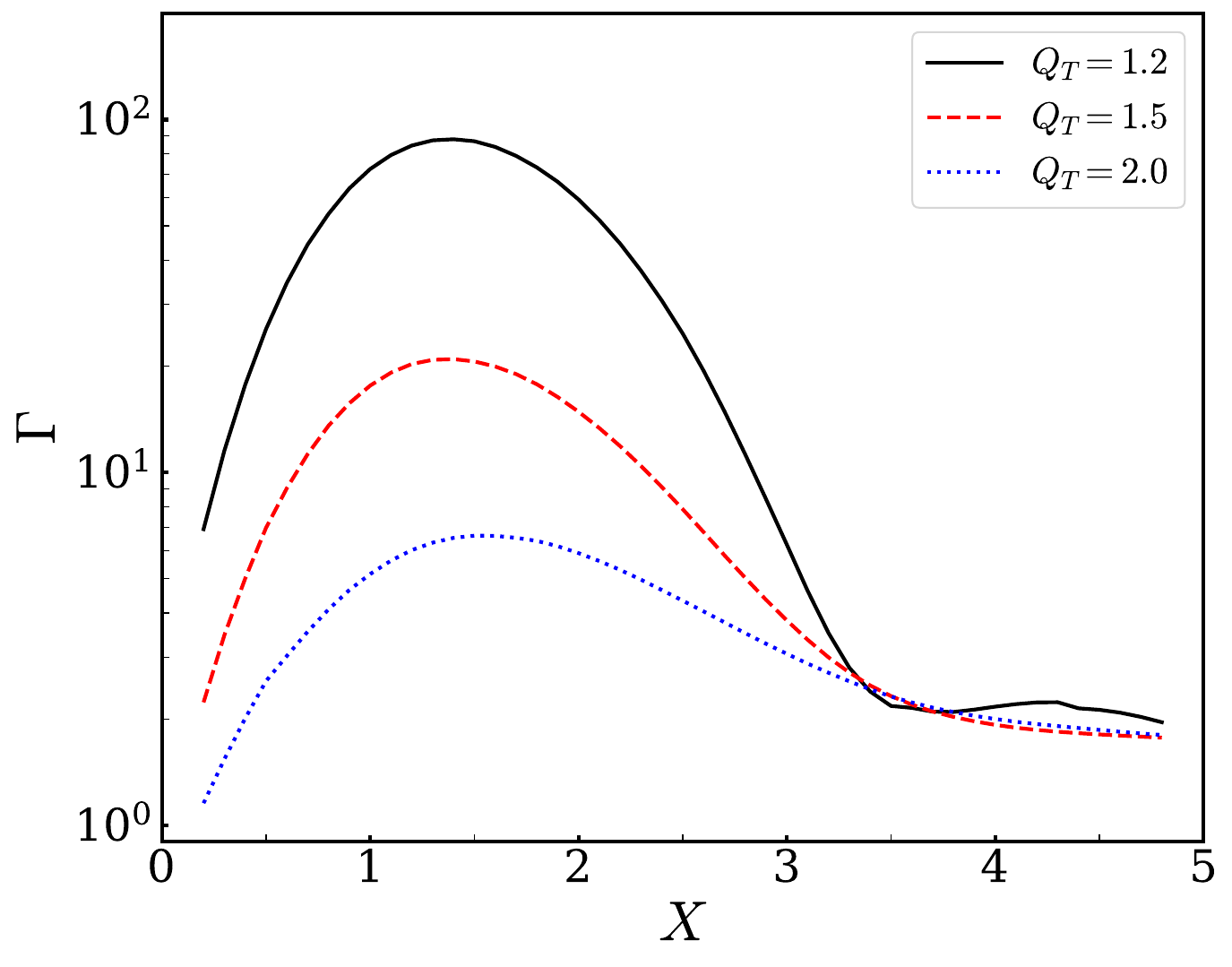}
\caption{Amplification factor $\Gamma$ as a function of $X$ for disks with $\QT =1.2$ (black solid), $\QT = 1.5$ (red dashed), and $\QT = 2.0$ (blue dotted).
\label{fig:ximax}}
\end{figure}

To quantify the magnitude of swing amplification, we define the amplification factor as $\Gamma \equiv \xi_\text{max}/\xi_\text{0}$, where $\xi_\text{max}$ denotes the maximum value of $|\xi|$ attained at $\tau > 0$. \cref{fig:ximax} plots $\Gamma$ as functions of $X$ for $Q_T=1.2$, $1.5$, and $2.0$. It is evident that $\Gamma$ increases with decreasing $Q_T$. The amplification factor reaches its maximum at $X \sim 1.4$, almost independent of $Q_T$, and decreases for both larger and smaller values of $X$, consistent with the previous results \citep[e.g.,][]{JT66,toomre81}. While self-gravity tends to enhance $\Gamma$, both random stellar motions and epicyclic motions act to decrease it. The stabilizing influence of random motions is most pronounced on small scales ($X \to 0$ or $k_y \to \infty$), as reflected in the decrease of the reduction factor $\mathcal{F}$ at large $x$. In addition, wavelets with large $k_y$ wind up rapidly, reducing the time available for the growth of small-scale perturbations \citep[see e.g.,][]{kim01}. The combined effect of these two processes leads to a decrease in $\Gamma$ as $X \to 0$. In contrast, at large $X$ (or small $k_y$), the gravitational term is diminished relative to the epicyclic term in \cref{eq:s}, thereby leading to a reduction in $\Gamma$.

\subsection{Comparison with Numerical Results}
\begin{figure}[t]
\centering
\epsscale{1.0} \plotone{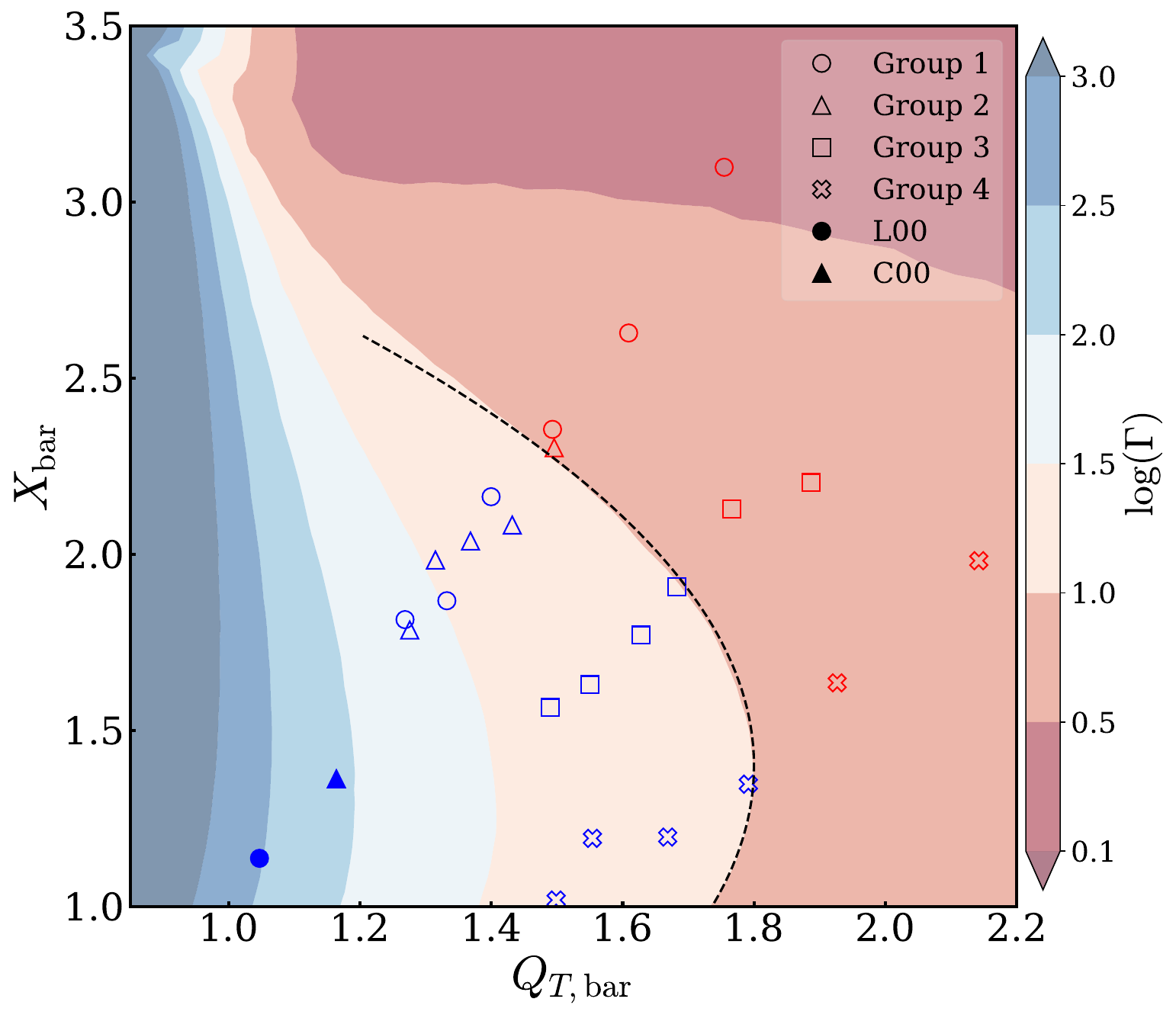}
\caption{Distribution of the amplification factor $\Gamma$ in the $\Qbar$--$\Xbar$ plane. Various symbols indicate the simulation outcomes: blue for bar-forming models and red for stable models. Open symbols represent models from this study, with circles, triangles, squares, and crosses corresponding to galaxies in Groups 1, 2, 3, and 4, respectively. The filled triangle and filled circle mark models \texttt{C00} and \texttt{L00} from \citet{jnk23}, respectively. The dashed line denotes the bar formation boundary corresponding to $\Gamma = 10$, as defined by the criterion in \cref{eq:QXfit}.
\label{fig:xqmap}}
\end{figure}

As discussed in \autoref{sec:barform}, bar formation in our models results from multiple cycles of swing amplification coupled with feedback mechanisms. To understand this process more quantitatively, we analyze it in terms of the amplification factor $\Gamma$ associated with a single swing amplification event. Since $\QT$ and $X$ vary with radius $R$, we compute the averaged values $\Qbar$ and $\Xbar$ over the radial range $2\kpc \leq R \leq R_{\QTmin}$ as representative values for each galaxy model. These values are listed in Columns (9) and (10) of \autoref{tbl:model}, respectively.
Owing to their lower stellar surface densities, galaxies in Group 1 tend to have larger values of $X$ than those in Group 4, despite also having lower $V_\text{max}$. 
\autoref{fig:xqmap} plots a contour map of the amplification factor $\Gamma$ in the $\Qbar$--$\Xbar$ parameter space. The positions of our model galaxies are overlaid as various open symbols: blue symbols represent models that develop bars, while red symbols correspond to those that remain stable. Circles, triangles, squares, and crosses indicate models belonging to Groups 1, 2, 3, and 4, respectively. For reference, two bulgeless models \texttt{C00} and \texttt{L00} from \citet{jnk23} are shown as filled triangle and filled circle symbols, respectively.

Note that all bar-forming models fall within the region characterized by an amplification factor of $\Gamma \gtrsim 10$. In contrast, models in the region with $\Gamma < 10$ undergo swing amplification that, even when sustained by repeated feedback loops, remains insufficient to trigger bar formation within $10\Gyr$. The region with $\Gamma \gtrsim 10$ is well approximated by 
\begin{equation}\label{eq:QXfit}
\Qbar + 0.4(\Xbar-1.4)^2\leq 1.8,
\end{equation}
whose boundary is delineated by the dashed curve in  \autoref{fig:xqmap}. For galaxies spanning a range of masses and sizes, $\Qbar \sim 1.8$ serves as the upper limit for bar formation. The critical value of $\Qbar$ decreases as $\Xbar$ deviates from 1.4, reflecting the diminished role of self-gravity or the effects of excessively rapid kinematic winding.

Finally, we remark on the stable models. \cref{fig:a2a0} shows that models with sufficiently small $a_h$ in each group remain at $A_2/A_0 < 0.2$ and therefore fail to develop a bar. Some of these models (e.g., \texttt{G1A31} and \texttt{G2A28}) reach $A_2/A_0 \sim 0.15$, forming only a weak central oval, whereas others (e.g., \texttt{G1A25} and \texttt{G4A35}) have $A_2/A_0 < 0.06$, corresponding to nearly featureless disks. The outcomes of these stable models can, in fact, be anticipated from their locations in the $\Qbar$--$\Xbar$ plane shown in \cref{fig:xqmap}, such that models with smaller $\Gamma$ yield correspondingly smaller values of $A_2/A_0$. This again demonstrates that swing amplification is the primary driver of perturbation growth in disk galaxies.

\section{Discussion} \label{sec:discussion}   

Here, we discuss our numerical results on bar formation in the context of the Ostriker–Peebles condition and the ELN criterion. We also compare the strengths and lengths of the simulated bars with observational data and investigate the conditions that lead to buckling instability.

\subsection{Bar Formation }\label{subsec:discuss2-1}

\begin{figure}[t]
\centering
\epsscale{1.0} \plotone{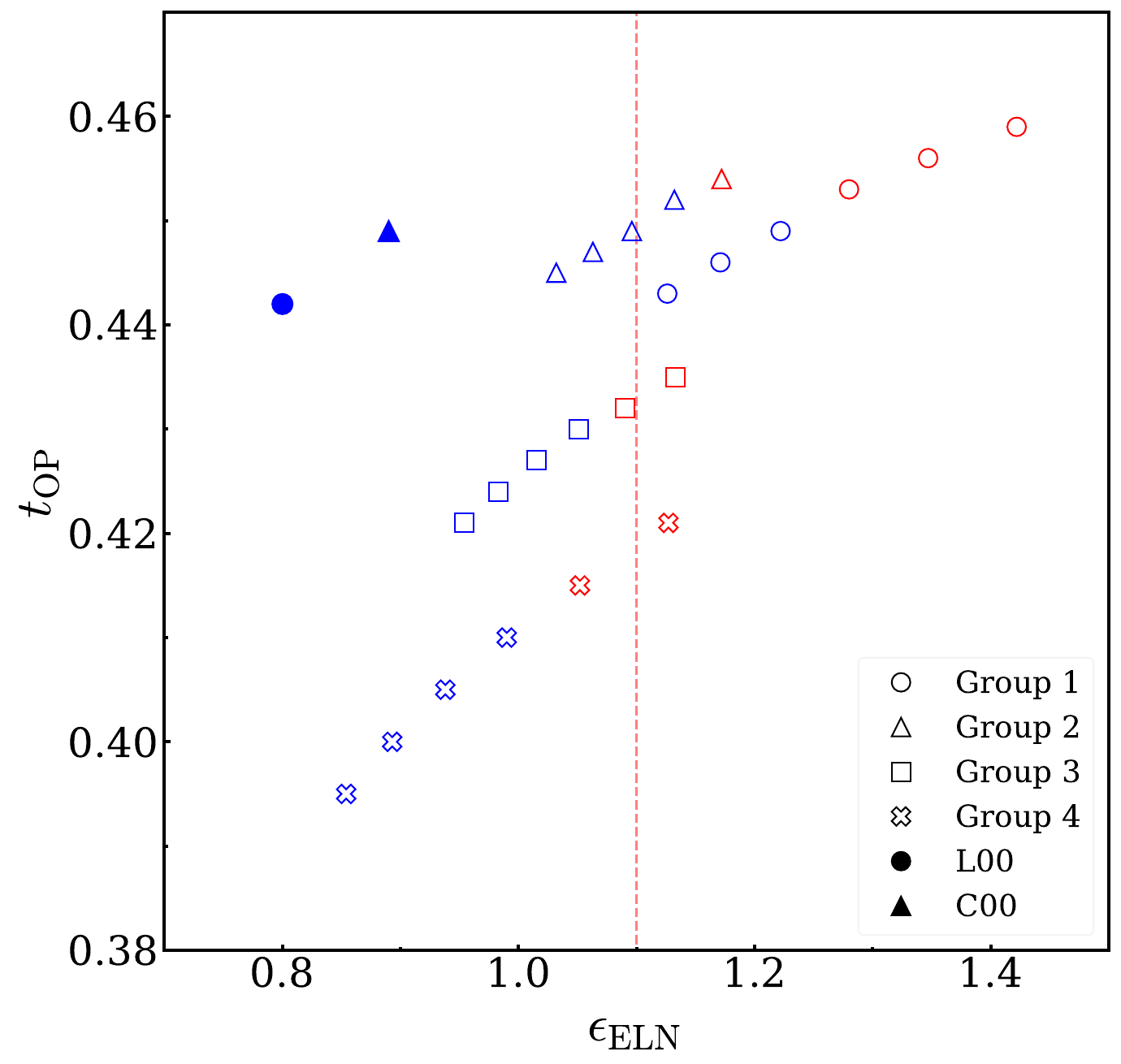}
\caption{Simulation outcomes in the $t_\text{OP}$--$\epsilon_\text{ELN}$ plane. The symbols have the same meanings as in \autoref{fig:xqmap}. The vertical dashed line marks $\epsilon_\text{ELN} = 1.1$ (see \autoref{e:eELN}).
\label{fig:op_eln}}
\end{figure}

Using bulgeless galaxy models, we have shown that systems satisfying the condition in \cref{eq:QXfit} become unstable to bar formation within $10\Gyr$. This two-parameter criterion, based on the swing amplification threshold of $\Gamma = 10$, effectively accounts for the outcomes of the $N$-body simulations. It is also of interest to examine whether the one-parameter conditions introduced in \autoref{sec:intro} can similarly reproduce the numerical results. \autoref{fig:op_eln} presents the simulation outcomes in the $t_\text{OP}$--$\epsilon_\text{ELN}$ plane, with blue and red symbols indicating models that are unstable and stable to bar formation, respectively. Although all of our models satisfy the Ostriker–Peebles condition for bar instability (\autoref{eq:top}), some remain stable without forming bars, indicating that $t_\text{OP}$ is not a reliable indicator of disk stability against bar formation \citep[see also][]{jnk23}. The ELN criterion, given in \cref{e:eELN}, also fails to accurately predict whether a galaxy will form a bar \citep[see also][]{yu15,algorry17,mario22,izq22}. We thus conclude that the two-parameter condition provides a more reliable and physically motivated criterion for predicting bar formation in bulgeless disk galaxies than the traditional one-parameter approaches.

Since our galaxy models without a bulge have $\text{CMC} \approx 0$, the criterion given in  \cref{e:cri}, which applies to Milky Way-like galaxies, appears to be more stringent than the condition given by \cref{eq:QXfit} for bulgeless galaxies. Indeed, two Milky Way-size models \texttt{C00} and \texttt{L00} from \citet{jnk23} lie in the region with $\Gamma>100$ in \autoref{fig:xqmap}, indicating that they form bar quite fast.  We note that \citet{jnk23} investigated the influence of a classical bulge and halo concentration by focusing on models with $\QTmin < 1.2$. The presence of a bulge alters the central rotation curve, leading to an increase in both the CMC and $\kappa$, which raises $X$ and enhances disk stability. In addition, the bulge modifies the radial profiles of $\QT$ and $X$. When these parameters vary rapidly with radius, the assumptions underlying the local analysis in \autoref{sec:amp} may no longer hold. It would therefore be worthwhile to investigate how the inclusion of a bulge alters the bar formation criterion across galaxies with different masses.

Our numerical results indicate that, within a given group, galaxies with lower $V_\text{max}$ tend to form bars earlier, whereas those with $V_\text{max}$ higher than the critical value remain stable and do not develop bars. The critical values of $V_\text{max}$ above which bar formation is suppressed are $\sim77$, 113, 170, and $250\ \kms$ for Groups 1, 2, 3, and 4, respectively. These thresholds lie below the mean $\overline{V}_\text{max}$ values for Groups 1 and 2, but exceed those for Groups 3 and 4, as reported in \autoref{sec:sample}. In fact, galaxies in Group 1 do not form bars even when $V_\text{max}$ falls almost $0.5 \sigma_{V_\text{max}}$ below the group mean. This suggests that low-mass disk galaxies, when modeled in isolation with only a stellar disk and dark halo, are generally less susceptible to bar formation than observed galaxies of comparable mass. Since real galaxies are subject to additional influences such as gas content, halo spin, and environmental interactions, the absence of bars in some models likely reflects limitations in the adopted initial conditions rather than a discrepancy with observations.

In particular, observations show that low-mass disk galaxies tend to have a high gas fraction, with $M_\text{gas}/M_d\sim 0.7$ \citep{diaz16,diaz20}. The role of the gaseous component in bar formation remains a subject of debate. According to \cite{beren07}, the timing of bar formation does not appear to be strongly correlated with the gas fraction. However, \cite{ath13} suggested that disks with higher gas content tend to resist bar formation, maintaining an axisymmetric structure for longer durations. \citet{seo19} argued that gas can promote bar formation in dynamically cold stellar disks, whereas in warm stellar disks, it tends to delay bar development. 
The presence of gas may even act to lock a bar by maintaining its pattern speed \citep{ath14,beane23}.
In addition, when strong turbulence is present in the gaseous component, $V_\text{max}$ inferred from \ion{H}{1} line widths may not solely reflect rotational motion, but may also include significant contributions from turbulent velocities. This suggests that low-mass galaxies may be slower rotators than implied by \autoref{fig:sample}, and that some galaxies classified as stable in our models may, in reality, lack sufficient rotational support to become bar-unstable once the contribution of turbulent motions is properly accounted for. Future work that includes a realistic gaseous component will be essential for more accurately assessing bar formation in low-mass disk galaxies.

Recently, \citet{ver24} investigated bar formation in galaxies spanning a range of masses, incorporating stellar and gaseous disks, a classical bulge, and a dark halo, constructed to resemble the properties of PHANGS-ALMA sample \citep{leroy21}.  They found that most of their models develop bars, with bar formation delayed by the presence of a bulge, while the inclusion of gas tends to promote bar formation. In contrast to our results, all of their bulgeless galaxies form bars. This discrepancy may arise from their use of an Einasto profile for the dark halo \citep{einasto65,ludlow17}, which has a significantly lower central density than the Hernquist profile adopted in our models. Consequently, their galaxies tend to exhibit lower values of $Q_T$ in the bar-forming region, rendering them more susceptible to bar instability. These results underscore the important role of the dark matter halo profile in determining bar stability.

\begin{figure}[t]
\centering
\epsscale{1.0} \plotone{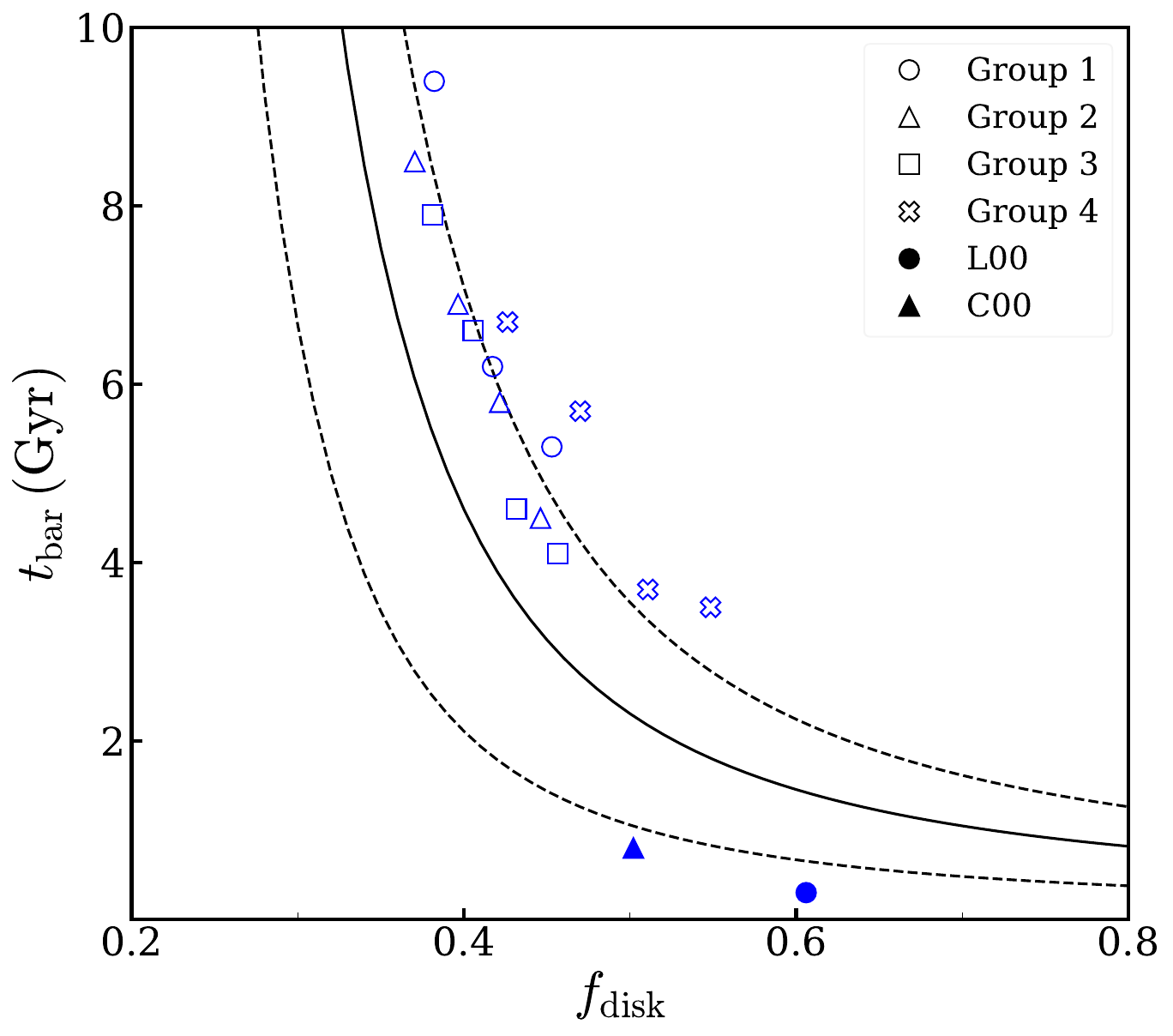}
\caption{Bar formation time $t_\text{bar}$ against the disk mass fraction $f_\text{disk}$ in our bar-forming models.  The symbols have the same meanings as in \autoref{fig:xqmap}. 
The solid line indicates the mean relation proposed by \citet{fujii18}, while the dashed lines represent the associated uncertainty bounds (see \autoref{eq:tbar}).
\label{fig:tb_fd}}
\end{figure}

\citet{fujii18} showed that the bar formation time $t_\text{bar}$ in their models follows the relation
\begin{equation}\label{eq:tbar}
t_\text{bar} = (0.146 \pm 0.079) \exp\left(\frac{1.38 \pm 0.17}{f_\text{disk}}\right) \Gyr,
\end{equation}
where $f_\text{disk}$ is the disk mass fraction, defined as the ratio of the disk mass to the total galaxy mass enclosed within $2.2R_d$. \autoref{fig:tb_fd} compares this relation with the bar formation times obtained from our simulations. The solid line indicates the mean relation, while the dashed lines represent the uncertainty range in the prefactor. Our numerical results are broadly consistent with the ``Fujii relation'', although they are systematically longer by $\sim 2\Gyr$. This discrepancy likely arises because all models in \citet{fujii18} adopt $\QT = 1.2$ at $R = 2.2R_d$, whereas our models have a higher average value of $\QT \sim 1.6$. In contrast, models \texttt{L00} and \texttt{COO} form bars significantly earlier than predicted by the Fujii relation, as their disks have a lower average value of $\QT \sim 1.0$ at $R = 2.2R_d$. We also note that variations in the halo mass distribution may contribute to differences in bar formation time.

Physically, the timing of bar formation is governed by two key factors: the swing amplification factor and the orbital timescale. The orbital timescale determines how rapidly swing amplification can proceed, while the amplification factor sets the number of amplification and feedback episodes required for bar formation. The strong correlation between $t_\text{bar}$ and $f_\text{disk}$ shown in \autoref{fig:tb_fd} suggests that both factors are effectively encapsulated by the disk mass fraction: the amplification factor depends primarily on the disk mass, while the orbital timescale is governed by the halo mass. 
Within a given group, galaxies with higher values of $\Gamma$ tend to form bars earlier, as one or two episodes of swing amplification are sufficient to generate a bar. Across groups, galaxies in Group 1 generally form bars later than those in Group 4, primarily due to their longer orbital times.

\begin{figure}[t]
\epsscale{1.0} \plotone{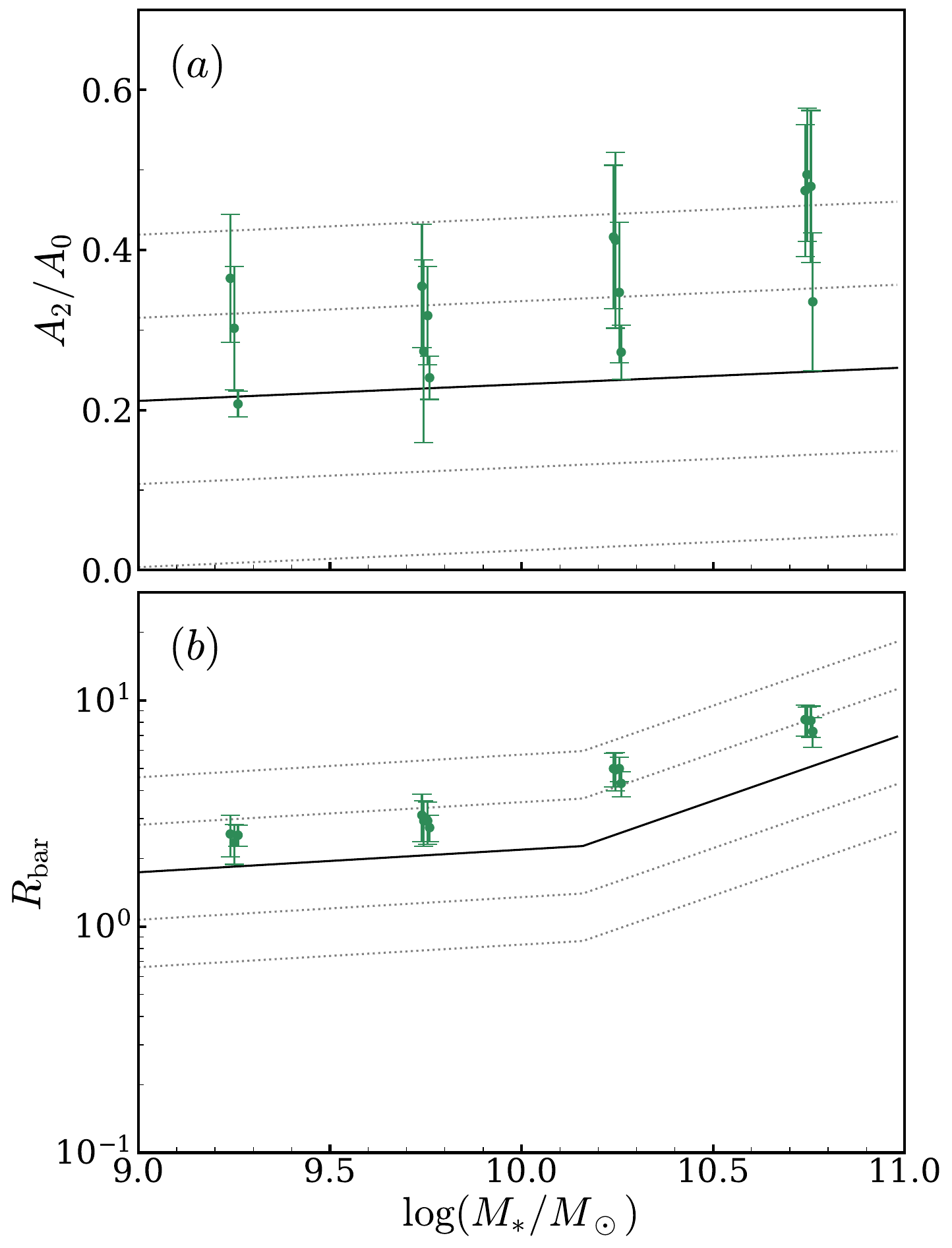}
\caption{Dependence on stellar mass $M_*$ of ($a$) bar strength $A_2/A_0$ and ($b$) bar length $R_\text{bar}$. Green dots and error bars represent the mean values and standard deviations from our bar-forming models. Bars within each group have identical stellar masses; data points are slightly offset along the horizontal axis for clarity. In panel ($a$), the solid line represents the linear fit to the observational data reported by \citet{lee22}, with dotted lines indicating the $1\sigma_{A_2/A_0}$ and $2\sigma_{A_2/A_0}$ deviations from the mean. In panel ($b$), the solid line represents the mean bar length for S$^4$G galaxies from \citet{erwin19}, with dotted lines indicating the $1\sigma_{R_\text{bar}}$ and $2\sigma_{R_\text{bar}}$ deviations.
\label{fig:RbarObs}}
\end{figure}
\subsection{Bar Properties}\label{subsec:discuss2-2}

Since our simulations cover galaxies with varying masses, it is of great interest to compare the numerical results with observations. Observational studies generally report no significant correlation between $A_2/A_0$ and host galaxy properties such as luminosity or Hubble type \citep{cuo20,kim21,lee22}, although a weak positive trend with circular velocity has been noted \citep{lee22}.  \autoref{fig:RbarObs}($a$) plots the relationship between bar strength $A_2/A_0$ and stellar mass $M_*$ in our models, compared to that derived from observations. The green dots and error bars represent the mean values and standard deviations of the bar strength, averaged over the period from bar formation to the end of the runs. In some models (e.g., \texttt{G1A34}, \texttt{G2A30}, \texttt{G3A29}, and \texttt{G4A45}), bars do not have sufficient time to develop fully, resulting in lower $A_2/A_0$ values compared to other models in their respective groups. Bars in Groups 1 and 2 are relatively weak, with $A_2/A_0 < 0.4$, whereas fully developed bars in Groups 3 and 4 are comparatively strong, with $A_2/A_0 > 0.4$. The solid line represents the linear fit to the observational data from \citet{lee22}, and the dotted lines indicate the $1\sigma_{A_2/A_0}$ and $2\sigma_{A_2/A_0}$ deviations from the mean. Although the bars in our simulations are stronger than those observed by less than the $2\sigma$ level, the gradual increase of $A_2/A_0$ with $M_*$ is consistent with the observational trend.

\begin{figure}[t]
\epsscale{1.0} \plotone{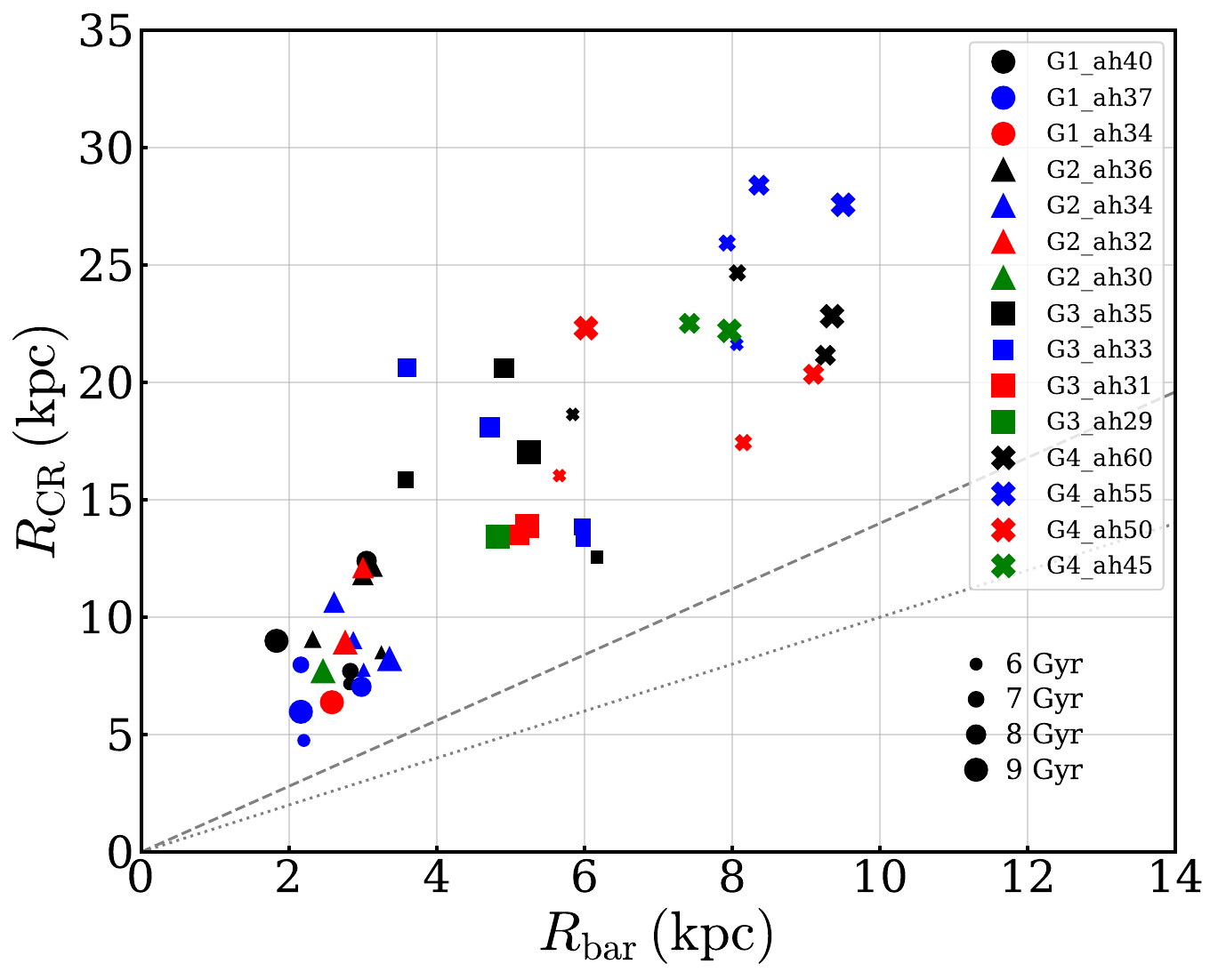}
\caption{Relationship between the corotation radius $R_\text{CR}$ and the bar length $R_\text{bar}$ for all bar-forming models in our simulations. The dashed and dotted lines correspond to $\mathcal{R} \equiv R_\text{CR}/R_\text{bar} = 1.4$ and 1.0, respectively. All bars satisfy $\mathcal{R} > 1.4$, indicating that they are classified as slow.
\label{fig:Ratio}}
\end{figure}

The bars in our simulations evolve in size over time, with higher-mass galaxies generally developing longer bars (see \autoref{fig:lbar}). \autoref{fig:RbarObs}($b$) compares the relationship between bar length $R_\text{bar}$ and stellar mass $M_*$ in our simulations with that in observations. The green dots and error bars again represent the mean values and standard deviations of the bar length in our simulations. The solid line draws the mean observational relation from the S$^4$G galaxy sample \citep{erwin19}, while the dotted lines indicate the $1\sigma_{R_\text{bar}}$ and $2\sigma_{R_\text{bar}}$ deviations from the mean.
The increasing trend of $R_\text{bar}$ with $M_*$ is broadly consistent between simulations and observations, although the simulated bars are typically longer by $\sim60\%$, corresponding to $\sim 1\sigma_{R_\text{bar}}$,  relative to the observed values.  The stronger and longer bars in our simulations are likely due to the absence of a classical bulge, which is known to weaken bars and limit their length \citep[e.g.,][]{sell80,se18,kd18,fujii18,jnk23}.

Bars are classified as slow or fast based on the ratio of the corotation radius $R_\text{CR}$ to the bar length $R_\text{bar}$. A bar is considered fast if $\mathcal{R} \equiv R_\text{CR}/R_\text{bar} < 1.4$ and slow if $\mathcal{R} > 1.4$. \autoref{fig:Ratio} plots the relationship between $R_\text{CR}$ and $R_\text{bar}$ for bars at various times throughout our simulations, indicating that they are all slow bars with $\mathcal{R}\sim(2.0-4.5)$.  While \citet{cuo20} reported that observed bars generally classified as fast, with a mean value of $\mathcal{R}\sim0.92$, bars in cosmological hydrodynamical simulations are predominantly slow, exhibiting mean values of $\mathcal{R} \sim 1.9$–$3.0$ \citep{roshan21}. \citet{frankel22} suggested that this discrepancy may arise from simulated bars being systematically shorter than those observed. In observational studies, the determination of $R_\text{CR}$ depends critically on the measurement of the bar pattern speed, which is subject to significant uncertainties \citep[e.g.,][]{jnk23}. In addition, the measured bar length depends on the adopted method. The values in \citet{erwin19} are based on the visual estimates of \citet{Her15}. Automated techniques for estimating bar length, such as the Fourier-based method \citep{lee20,lee22} and ellipse fitting \citep{erwin03,lauri04,agu09}, are generally known to underestimate bar length. However, when spiral arms or rings are present, ellipse fitting can instead overestimate bar length \citep{cuo21,Lee25}. \cite{petersen24} also reported that ellipse fitting often yields bar lengths exceeding the dynamical length inferred from the radial extent of trapped orbits.

\subsection{Buckling Instability}\label{subsec:discuss3}

In this study, we have shown that only massive galaxies undergo buckling instability, whereas low-mass galaxies remain vertically thin throughout their evolution. The ratio $\sigma_z/\sigma_R$ of the velocity dispersion has frequently been considered a key diagnostic for identifying the onset of buckling instability in galactic bars, as it reflects the anisotropy in stellar kinematics that can trigger vertical instabilities. In our simulations, we find that the timing of buckling in Groups 3 and 4 closely corresponds to the epoch when $\sigma_z/\sigma_R$ drops below approximately 0.6 at a representative radius of $R = 2\kpc$. This threshold appears to be consistent with previous studies that suggest a critical value of $\sigma_z/\sigma_R$ below which bars become susceptible to bending modes. However, this criterion does not universally predict buckling: galaxies in Groups 1 and 2 exhibit $\sigma_z/\sigma_R$ values below 0.55 without undergoing buckling, indicating that a low $\sigma_z/\sigma_R$ alone is not a sufficient condition. This suggests that additional factors, such as bar strength, vertical scale height, the presence of outer spirals, the degree of a CMC, and the dynamical maturity of the bar, may play a significant role in determining whether a barred galaxy experiences buckling instability.

In our previous work, \citet{jnk23} showed that only bulgeless galaxies undergo buckling instability, while galaxies with a bulge remain stable due to the stabilizing influence of the CMC, which suppresses vertical bending modes and enhances the vertical restoring force. More recent simulations have demonstrated that buckling can still occur in galaxies with a central bulge \citep{li24,mcclure25}, although the bulges in those models are significantly less massive than the disks, resulting in a relatively low CMC that is insufficient to fully suppress vertical instabilities. In addition, halo spin has also been shown to influence the onset and strength of buckling instability \citep{jnk24}. These findings collectively indicate that buckling is governed by a complex interplay of many factors, including bulge mass, halo spin, bar strength, and disk structure, and thus caution must be exercised when interpreting $\sigma_z/\sigma_R$ or any single parameter as a universal criterion for buckling \citep[e.g.,][]{li23a,li23b}.

\section{Summary} \label{sec:summary}

We have conducted a series of $N$-body simulations of bulgeless disk galaxies to investigate the conditions under which bars form, focusing on systems with stellar masses ranging from $10^9$ to $10^{11}\Msun$. To facilitate a systematic analysis, we divide the models into four groups based on their stellar mass. For each group, the galaxy models are constructed to match the observed properties of nearby barred galaxies from the S$^4$G survey. To explore the influence of the halo concentration on bar formation, we vary the halo scale radius systematically, while keeping other structural parameters fixed. This approach allows us to isolate the dynamical role of the rotational velocity in driving or suppressing bar instabilities across the mass spectrum. Our main findings are summarized as follows:

\begin{enumerate}

\item \emph{Bar formation} -- We demonstrate that bars in our models form through repeated swing amplification of density waves intrinsic to the particle distribution in a stellar disk, reinforced by feedback loops. A strong \ac{ILR} suppresses direct wave feedback across the galactic center, implying that the regeneration of leading waves from trailing ones likely proceeds through a combination of reflections at the \ac{ILR} and nonlinear interactions among waves with differing wavenumbers. These processes collectively sustain the amplification cycle necessary for bar formation.

\item \emph{Conditions for bar formation} -- Linear theory of swing amplification predicts that the amplification factor $\Gamma$ depends on two parameters: the Toomre stability parameter $\QT$ and the dimensionless wavelength $X$, increasing as $\QT$ decreases and peaking near $X \sim 1.4$. Using the radially averaged quantities $\Qbar$ and $\Xbar$ over the range $2\kpc < R < R_{\QTmin}$, we find that all galaxy models that develop a bar fall within the region where $\Gamma \gtrsim 10$, as characterized by \cref{eq:QXfit}. We therefore propose this relation as a two-parameter criterion for bar formation in disk galaxies without a bulge. Traditional one-parameter conditions, such as the Ostriker–Peebles and ELN criteria, fail to account for the bar formation observed in our numerical simulations.

\item \emph{Bar Evolution} -- Our simulations reveal a clear dependence of bar evolution on the galaxy mass. In low-mass systems (Groups 1 and 2), bars that form are relatively short and weak, and highly susceptible to destructive interference from outer spiral arms. In contrast, more massive galaxies (Groups 3 and 4) develop longer and stronger bars that form earlier and are less affected by spiral arms, owing to their higher self-gravity and weaker outer spirals. 

\item \emph{Bar properties} -- The strengths and lengths of the bars formed in our simulations are broadly consistent with observational trends. The bar strength $A_2/A_0$ increases gradually with stellar mass $M_*$, in line with observations, although the simulated bars are generally stronger, falling within approximately $2\sigma_{A_2/A_0}$ of the observed values. Similarly, the bar length $R_\text{bar}$ in our simulations is systematically greater than that of observed galaxies of comparable mass, exceeding the mean observed values by $\sim1\sigma_{R_\text{bar}}$. These discrepancies may be attributed to the absence of a classical bulge in our models, as classical bulges are known to suppress bar formation and limit bar length. Bar pattern speeds in the simulations decline over time due to angular momentum transfer from the bar to the surrounding dark matter halo. As a result, all simulated bars are classified as slow, with corotation-to-bar length ratios in the range $\mathcal{R} = R_\text{CR}/R_\text{bar} \sim 2.0$–$4.5$.

\item \emph{Buckling instability} -- In low-mass galaxies (Groups 1 and 2), bars remain vertically thin and do not undergo buckling instability, even when the ratio $\sigma_z/\sigma_R$ drops below 0.55. In contrast, bars in more massive galaxies (Groups 3 and 4) do undergo buckling when $\sigma_z/\sigma_R$ drops below approximately $0.6$. This indicates that a low value of $\sigma_z/\sigma_R$ alone is not a sufficient condition for the onset of buckling. Additional factors including bar strength, vertical scale height, the presence of outer spiral arms, and the degree of CMC are likely to play a role in triggering the instability.

\end{enumerate}

\section*{acknowledgments}
We are grateful to the referee for providing a constructive and thoughtful report.
The work of D.J.\ was supported by Basic Science Research Program through the National Research Foundation of Korea (NRF) funded by the Ministry of Education (RS-2023-00273275). The work of W.-T.~K.\ was supported by the grant of the National Research Foundation of Korea (RS-2025-00517264).  YHL acknowledges support from the Basic Science Research Program through the National Research Foundation of Korea (NRF) funded by the Ministry of Education (RS-2023-00249435). Computational resources for this project were provided by the Supercomputing Center/Korea Institute of Science and Technology Information with supercomputing resources including technical support (KSC-2024-CRE-0532).

\bibliography{1ms}

\end{document}